# Electronic structure, spectroscopy, cold ion-atom elastic collision properties and photoassociation formation prediction of $(MgCs)^+$ molecular ion

Mohamed Farjallah[†a], DibyenduSardar[†b], Bimalendu Deb[b], and Hamid Berriche[a,c]

[a]Laboratory of Interfaces and Advanced Materials, Faculty of Sciences of Monastir, University of Monastir, 5019 Monastir, Tunisia

[b]School of Physical Sciences, Indian Association for the Cultivation of Science (IACS), Jadavpur, Kolkata 700032, INDIA

[c]Department of Mathematics and Natural Sciences, School of Arts and Sciences, American University of Ras Al Khaimah, RAK, P. O. Box 10021, UAE

†Author Contributions: M.F. and D.S. contributed equally to this manuscript.

## Abstract

Studies on the interactions between an alkali atom and an alkaline earth ion at low energy are important in the field of cold chemistry. In this paper we, extensively, study the structure, interactions, and dynamics of $(MgCs)^+$ molecular ion. We use an ab initio approach based on the formalism of non-empirical pseudo-potential for $Mg^{2+}$ and $Cs^+$ cores, large Gaussian basis sets and full valence configuration interaction. In this context, the $(MgCs)^+$ cation is treated as an effective two-electron system. Potential energy curves and their spectroscopic constants for the ground and the first 41 excited states of $^{1,3}\Sigma^+$, $^{1,3}\Pi$ and $^{1,3}\Delta$ symmetries are determined. Furthermore, we identify the avoided crossings between the electronic states of $^{1,3}\Sigma^+$ and $^{1,3}\Pi$ symmetries. These crossings are related to the charge transfer process between the two ionic limits $Mg/Cs^+$ and $Mg^+/Cs$. In addition, vibrational-level spacings, the transition and permanent dipole moments are presented and analysed. Using the produced potential energy data, the ground-state scattering wave functions and elastic cross sections for a wide range of energies are performed. Furthermore, we predict the formation of translationally and rotationally cold molecular ion $(MgCs)^+$ in the ground state electronic potential energy by stimulated Raman type process aided by ion-atom cold collision. In the low energy limit (< 1 mK), elastic scattering cross sections exhibit Wigner law threshold behaviour; while in the high energy limit the cross sections as a function of energy $E$ go as $E^{-1/3}$. A qualitative discussion about the possibilities of forming the cold $(MgCs)^+$ molecular ions by photoassociative spectroscopy is presented.

## I. Introduction

In the recent years, cold and ultra-cold molecules [1] have received considerable attention owing to successful demonstrations of cooling and trapping of molecules [2] and their importance in high-resolution spectroscopy [3]. Because of the characteristic of the long-range ion–atom potential, cold ion–neutral collisions [4-6] are different from neutral–neutral and ion–ion collisions. The long range ion–atom potential is described by $-C_4/r^4$, where $C_4 = \alpha q^2/(8\pi\varepsilon_0)$, with α being the static electric polarizability of the atom and r the ion–atom separation. The scattering by this potential offers the primary channel for the exchange of energy between the ions and atoms. Accordingly, the investigations of ion–neutral atom collisions at low energy are very useful for the study of sympathetic cooling of atomic ion translational motion [7,8], and to get a detailed control of the internal and external degrees of freedom of the molecular ions [9]. In addition, interaction between the alkali atoms and alkaline earth ions ($Be^+$, $Mg^+$, $Sr^+$ and $Ca^+$) offer opportunities for new developments in the field of ultra-cold quantum matter, with a benefit of simpler and more reliable trapping compared to the neutral molecules. Furthermore, a wealth of cold molecular ion species could be created, opening the way to a rich chemistry at temperatures of a few mK or less [10]. Interactions between cold atoms and ions are also relevant for important applications related to molecule formation in Bose–Einstein condensates [11] and to quantum information. Photoassociation [12] and Feshbach resonance tuning [13] are two main experimental techniques for a coherent production of ultracold molecules from ultracold atoms.

Radiative emission during cold collisions between trapped laser-cooled Rb atoms and alkaline-earth ions $Ca^+$, $Sr^+$ and $Ba^+$, were studied theoretically by Aymar et al [14] using effective core potential based on quantum chemistry calculations of potential energy curves and transition dipole moments of the related molecular ions. Furthermore, many experimental groups have carried out experiments with various combinations of alkali atoms Rb and alkaline-earth atomic ions: Rb atoms with $Ca^+$ [15, 16] and $Ba^+$ [17] ions. Moreover, recently, other groups have studied the energy scaling of cold atom-atom-ion three-body recombination $Ba^+$+Rb+Rb in the mK regime where a single $^{138}Ba^+$ ion in a Paul trap is immersed into a cloud of ultracold $^{87}Rb$ atoms [18], and life and death of a cold $BaRb^+$ molecule inside an ultracold cloud of Rb atoms [19]. In addition, an optical dipole trap of Rb atoms has also been merged in a Paul trap containing a few $Ba^+$ atoms [20, 21].

Concerning $MgH^+$, the first evidence of the formation of molecular $MgH^+$ ions in a laser cooled ion trap was reported by Baba and Waki [22], by introducing air into an $Mg^+$ trap. Drewsen and colleagues [23], who introduced a thermal gas $H_2$or $D_2$ into a laser cooled $Mg^+$ trap creating trapped $MgH^+$ or $MgD^+$ ions, carried out a more controlled experiment. Many other studies concerning magnesium hydride investigated its photodissociation [24], branching ratios [25], etc. . More recently, Aymar et al. [26] carried out a study of the electron structure of the $MgH^+$ ionic molecule containing potential energy curves, transition and permanent dipole moment spectra and static polarizabilities. Using an ab initio approach, Khemeri et al [27] performed a study of the electronic properties of this molecule in the adiabatic representation.

About the $MgLi^+$ molecule, Boldyrev et al. [28] determined its ground state spectroscopic constants, using a split-valence basis MP2/6-311$^+$G*. Pyykkö [29] used two methods: Hartree-Fock (HF/6-31G *) and Møller-Plesset (MP) to calculate the equilibrium distance, well depth and vibration frequency for the same states. More recently, Yufeng Gao and Tao Gao [30] used the ab initio program package MOLPRO to determine the potential energy curves and the permanent and transition dipole moments of the same molecules. They used the multi-reference configuration interaction and valence full configuration interaction with large aug-cc-pCVQZ basis sets taking into account the core–valence and scalar relativistic corrections.

A review of the scientific literature shows that experimental studies on $(MgCs)^+$[31] cationic molecule are practically absent. We noticed the existence of a recent single theoretical study of $(MgCs)^+$, where the authors Smialkowski and Tomza[32] reported only spectroscopic constants for the ground state. Therefore, a more detailed and refined investigation of this molecular system is desirable. Methods fundamentally similar to those used for the treatment of $MgLi^+$ molecule [33, 34] will be employed in the present study.

Several groups in the world have experimentally investigated atom-ion and cold molecular diatomic mixtures. The list include $Yb_2^+$ [35], $RbCa^+$ [36],$CaBa^+$ [37], $CaYb^+$ [38], $RbYb^+$ [39], $LiCa^+$ [40], $RbCa^+$ [41], $NaCa^+$ [42], $RbSr^+$ [43], $LiYb^+$ [44],$KCa^+$ [45], $RbBa^+$ [46], $Rb_2^+$ [47-49], $BaRb^+$ [50], and $Ca_2^+$ [51]. Alkaline-earth ions are a common choice as advanced methods for manipulation and detection of such ions have been developed over the years, opening prospects for their applications in quantum simulation and computations[52-53]. Atom-ion diatomic molecular systems can be produced via cold collision-induced charge-transfer radiative association, light-induced photoassociation [54-58]or photoionization of ultra-cold neutral molecules [49,51]. The effective core potential methods followed by CI and quantum chemistry calculations are employed to calculate the potential energy curves and dipole moments. In a very recent work [59] in our group, we investigated the electronic structure of the five diatomic molecular ions composed of a $Ca^+$ ion interacting with an alkali-metal atom: $CaX^+$ (X=Li, Na, K, Rb, Cs).

Very recently, Smialkowski and Tomza [32] theoretically investigated the ground state electronic structure of single-charged molecular ions formed from two or three interacting alkali-metal and alkaline-earth-metal atoms. They calculated ground-state electronic properties of all diatomic $AB^+$ and most of triatomic $A_2B^+$ molecular ions consisting of Li, Na, K, Rb, Cs, Mg, Ca, Sr, Ba and Yb atoms. Smialkowski and Tomza [32] carried out a systematic study of the electronic structure of $MgX^+$ (X=Li, Na, K, Rb and Cs) ionic systems and presented the potential energy curves and spectroscopic constants for their ground states.

Here we investigate the adiabatic potential energy curves for many electronic states of $^{1,3}\Sigma^+$, $^{1,3}\Pi$ and $^{1,3}\Delta$ symmetries below the $Mg^{2+}Cs^-$ asymptotic limit. Then, we extract from these curves the spectroscopic constants (equilibrium distance $R_e$, well depth $D_e$, electronic excitation energy $T_e$, frequency $\omega_e$, harmonicity constant $\omega_e x_e$, and rotational constant $B_e$). In addition, vibrational-level spacings and the dipole moments function are also presented and analyzed. By exploiting the ab-initio potential energy data, we study elastic collisions between an alkali ion $Cs^+$ and a neutral alkaline earth atom Mg for the ground state collisional threshold. Besides that, we predict and analyze the formation a of translationally and rotationally cold molecular ion by two-photon incoherent Raman process in the presence of two applied lasers.

This paper is organized as follow. In section 2, the used methodology is briefly presented. We report our results for $(MgCs)^+$ and discuss them in Section 3. Finally, we conclude in section4.

## II. Theory and Computational Details

In this section, we provide an insight into the technical details of the computational calculation of the $(MgCs)^+$ molecular ion. Numerous studies on heteronuclear alkaline-earth dimers such as $(MgK)^+$ [56], $(BeLi)^+$ [60], $(BeH)^+$ [61], $(CaRb)^+$, $(SrRb)^+$ and $(BaRb)^+$ [41] have been performed. Using the CIPSI package (Configuration Interaction by Perturbation of a multi-configuration wave function Selected iteratively) of the Laboratoire de Physique et Chimie Quantique of Toulouse in France [62]. The ionic system $(MgCs)^+$ has 67 electrons. Therefore, we replaced the cores of the $Mg^{2+}$ (10 electrons) and $Cs^+$ (54 electrons) with non- empirical pseudo-potentials proposed by Barthelat and Durand [63-65]. Consequently, $(MgCs)^+$ is treated as an effective two-electron system, where the two valence electrons are moving in the field of the two cores. The pseudo-potential is complemented by the corrections of core-core and core-valence correlation according to the operator formalism of Muller et al [66]. In addition, the cut-off function reported by the formulation of Foucrault et al. [67] is taken to be a function of *l*, the orbital angular momentum, in order to treat separately the interaction of valence electrons of different spatial symmetry and the core electrons. The cut-off radii for the lowest valence s, p, d, and f one-electron states are reported in table 1. In the present work, the core polarizabilities are taken to be =0.46904$a_0^3$ [67] for $Mg^{2+}$ and 15.117$a_0^3$ [68-69], respectively. For the magnesium, we used a large Gaussian basis set (9s, 7p, 5d, and 4f) composed of 83 functions. The diffuse orbital exponents have been optimized to reproduce with a good accuracy all atomic states dissociating into: $Mg^+$(3s, 3p, 4s, 3d, 4p, 5s, 4d, 4f, 5p, and 6s) and Mg (3s$^2$ ($^1$S), 3s3p ($^3$P), 3s3p ($^1$P), 3s4s ($^3$S), 3s4s ($^1$S), 3s3d ($^1$D), 3s4p ($^3$P), 3s3d ($^3$D), 3s4p ($^1$P), 3s5s ($^3$S), 3s5s ($^1$S), 3s4d ($^1$D), 3s4d ($^3$D), and 3s5p ($^3$P)). After contraction, this basis was reduced to 7s/7p/4d/4f and the function number decreases to only 76. Aymar et al. [68]used a (7s/5p/4d/2f) basis set contracted to (6s/5p/2d/2f); therefore, the function number decreases from 56 to 45, while for Cs, we used a (7s4p5d1f/6s4p4d) Gaussian basis set taken from [70-71].

Molecular energies for $(MgCs)^+$, at the dissociation limits, are given in Table 2. The quality of the used basis sets and cutoff radii are confirmed by the good agreement between our values and the experimental [72] and theoretical [73-74] atomic energy levels. The difference between our results and the experimental values for almost all energy levels is lower than 179 cm$^{-1}$, which is found for Mg (3s4s) atomic level. However, a difference of 1445 cm$^{-1}$ is found for Mg (3s3d) atomic level. Nevertheless, the agreement between the energy levels of $Mg^+$ and the experimental ones is much better seeing that we have reproduced $Mg^+$ experimental binding energies exactly for: 3s, 3p, 3d, and 4p. The difference with experimental values for $Mg^+$ (4s) atomic level is equal to 1.53 cm$^{-1}$ and for the other highly excited states 5s, 4d, 4f, 5p, and 6s, the discrepancy is between 22 and 105 cm$^{-1}$.

## III. Results and discussion

### III. 1. Adiabatic potential energy and spectroscopic constants

Using the method reported in the previous section, a full and expanded calculation were performed for many electronic states of $^{1,3}\Sigma^+$, $^{1,3}\Pi$ and $^{1,3}\Delta$ symmetries for the $(MgCs)^+$ cationic molecule. These states dissociate into $Mg^+$ (3s, 3p, 4s, 3d, 4p, 5s, 4d, 4f, 5p, and 6s) + Cs(*6s, 6p, 4d, 5d,7s and 7p*) and Mg {(3s$^2$ ($^1$S), 3s3p ($^3$P), 3s3p ($^1$P), 3s4s ($^3$S), 3s4s ($^1$S), 3s3d ($^1$D),

3s4p ($^3$P), 3s3d ($^3$D), ($^1$P), 3s5s ($^3$S), 3s5s ($^1$S), 3s4d ($^1$D), 3s4d ($^3$D), and 3s5p ($^3$P))} + $Cs^+$. The adiabatic potential energy is performed for an interval of intermolecular distances from 3.50 to 200.00 $a_0$ with a step of 0.1 $a_0$ around the avoided crossings. The potential energy curves as functions of the internuclear distance R for the $^{1,3}\Sigma^+$, $^{1,3}\Pi$ and $^{1,3}\Delta$ electronic states are drawn respectively, in figure 1 to figure 5.

In order to assess the precision of our calculated potential energy curves for the states previously studied we have extracted the spectroscopic constants(equilibrium distance $R_e$, well depth $D_e$, electronic excitation energy $T_e$, frequency $\omega_e$, harmonicity constant $\omega_e x_e$, and rotational constant $B_e$ for the equilibrium separation $R_e$). These constants were determined by least-squares interpolation of the vibrational energy levels.

Unfortunately, no experimental spectroscopic information for ($MgCs^+$) have been published yet. However, Smialkowski and Tomza. [32] recently, reported a theoretical calculation for the ground state. They found $R_e$=7.85 $a_0$, $D_e$= 1861 $cm^{-1}$, $\omega_e$=73.2$cm^{-1}$, and $B_e$= 0.0481 $cm^{-1}$, which are close to our values ($R_e$= 7.70 $a_0$, $D_e$= 2047 $cm^{-1}$, $\omega_e$=73.2$cm^{-1}$, and $B_e$= 0.049364 $cm^{-1}$). Therefore, the spectroscopic constants for the excited states are presented here for the first time. Consequently, we think that these potential energy and spectroscopic data can initiate future experimental and theoretical investigations on this molecular ionic system.

In Table 3, we report the spectroscopic constants for the ground and the excited states for all symmetries. The potential energy curves for the 20 first $^{1,3}\Sigma^+$ states of $(MgCs)^+$ molecular ion are plotted in figures 1 and 2. These states are labelled by increasing numbers from 1 to 10 for both singlet and triplet states. The ground ($X^1\Sigma^+$) and the first excited ($2^1\Sigma^+$) states are well separated from the highest states of the same symmetry and they maintain an energy gap of, about, 42012$cm^{-1}$. Both are found with unique minimums at 7.7 and 11.45 $a_0$ and well depths of 2047 and 2662 $cm^{-1}$, respectively. The three first excited triplet states (1-3)$^3\Sigma$ exhibit also unique minimums at 8.58, 14.98 and 17.20 $a_0$ with significant well depths of 4504, 813 and 2658 $cm^{-1}$, respectively.

An accurate analysis of these curves shows that the most strongly bound states are the first excited sate of both singlet and triplet states with well depths of 2662 and 4504 $cm^{-1}$ located at 11.45 and 8.58 $a_0$, respectively.

Concerning $^{1,3}\Pi$ symmetry, we have drawn 16 PECs for both multiplicities. These states are presented in figures 3-4 as a function of separations R. Based on those curves and their spectroscopic constants; we note that the $^{1,3}\Pi$ states are weakly bound with well depths that do not exceed 1200 $cm^{-1}$.As for the $^{1,3}\Delta$ symmetries, the first six PECs are plotted in figure 5as. These states are labelled by increasing numbers:1, 2 and 3 for each multiplicity. We found that all these states are repulsive. Therefore, this cationic molecular system is unstable in the 1-$3^{1,3}\Delta$ states.

The analysis of all these curves shows that the singlet and triplet potential energy curves of $\Sigma^+$, $\Pi$ and $\Delta$ symmetries tend to the same asymptotic limits. For example, the ($1^{1,3}\Delta$) potential energy curves dissociate into the same limit Mg(3s3p)+$Cs^+$, leading to degenerated singlet and triplet states at the dissociation limit.

An interesting feature of the higher excited states of all symmetries of $\Sigma^+$ and $\Pi$ states has been noticed. The shape becomes more involved, with double wells and potential barriers. This feature is due to the presence of many avoided crossings at short and large values of internuclear distances between excited states of same symmetries. The positions of such avoided crossings are presented in figures 1 and 2. These series of avoided crossings are related to underlying charge transfer between the ionic configurations $Mg^+$/Cs and Mg/$Cs^+$. Later, these avoided crossings will be discussed in the light of the behaviour of the permanent dipole moments.

The PECs of excited states starting from $4^1\Sigma^+$ present double wells and barriers, which are the consequence of avoiding crossings that happen at specific inter-nuclear distances. These avoided crossings are related to underlying charge transfer as it can be seen from the change of location of the positive charge at dissociation limits between two neighbour states. The positive charge moves from an atom to the second one. The existence of these avoided crossings will generate large nondiabetic couplings. This same behaviour is seen for other symmetries such as $^3\Sigma^+$ states. However, for the $^3\Pi$ states multiple avoiding crossings are happening between and not only due to charge transfer as the positive charge keep located on the same atom for the two successive states that have shown such avoided crossings. We expect that in the diabatic representation the $4^3\Pi$ state should cross all higher excited states to provide reel crossings with these dissociating states.

### III.2. Vibrational levels

For a better understanding of the potential energy curves structure, the vibrational levels for all attractive states are computed using the Numerov algorithm. The interpolation of the potential curves is carried out with 12000 points and a grid ranging from 3.5 to 180.0 $a_0$.

The vibrational spacings ($E_{v+1}-E_v$) as a function of the vibrational quantum number v are reported in figures 6-7 for the $^{1,3}\Sigma^+$ states. As usual, these spacings are not constant due to the important anharmonicity observed in the potential energy curves.

The number of the vibration levels for the ground and the first excited states are 80 and 180, respectively. The vibrational energy spacing of these two potentials shows a linear behaviour for low vibrational levels, and then it vanishes at the dissociation limits.

The spacing between vibrational levels decreases gradually as vibrational energy increases. This behaviour is particularly prominent for vibrational levels of the PECs that reflect the strong anharmonicity. The overall pattern of energy spacings of different electronic states of the molecular ion is similar. However, for some states irregularities related to the avoided crossings are visible, e.g. for the $^1\Sigma^+$ states.

### III. 3. Permanent and Transition Dipole Moments

In addition to the potential energy curves, the practical implementation of cold molecule formation via photoassociation requires knowledge of their electronic properties such as the radial variation of permanent or transition dipole moments. Furthermore, the knowledge of the dipole function of molecular systems can be considered as a sensitive test for the accuracy of the calculated electronic wave functions and energies.

Keeping this broad perspective of the ion-atom systems in mind, we have determined the permanent and transition dipole moments for the same large and dense grid of internuclear distances used for the potential energy curves.

### III. 3.1 Permanent Dipole Moments.

The Permanent and Transition dipole moments are determined by considering the origin at atom Mg, and where the distance R between $Mg/Mg^+$ and $Cs/Cs^+$ stands for internuclear distance between them. We calculated the permanent dipole moments as a function of R, the intermolecular distance, for all studied states (1-10$^1\Sigma^+$ , 1-10$^3\Sigma^+$ , 1-8$^1\Pi$, 1-8$^3\Pi$, 1-3$^1\Delta$, and 1-3$^3\Delta$) are displayed in figures 8-12 .

At short and intermediate internuclear distance, the permanent dipole moments present undulations with abrupt variation between neighbour electronic states. For example, there is an abrupt variation between the $8^1\Sigma^+$ and $9^1\Sigma^+$ states located at 33.5 $a_0$. We remark that this particular distance represents approximately the position of the avoided crossing between the two potential energy curves that we noticed before. We can conclude that the significant changes of permanent dipole moment at small internuclear distance are due to the change of the polarity in the molecule. In addition, the positions of the irregularities in the R-dependence of the permanent dipole are correlated to the avoided crossings between potential energy curves, which are both manifestation of abrupt changes in the character of the electronic wave-functions.

For the large internuclear distance, the permanent dipole moment of the $X^1\Sigma^+$, $3^1\Sigma^+$, $5^1\Sigma^+$, $7^1\Sigma^+$ and $9^1\Sigma^+$ states, dissociating into Mg ($3s^2$, 3s3p, 3s3p, 3s4s, 3s4s, 3s3d, 3s4p, 3s3d, 3s5s, 3s5s, 3s4d, 3s4d, and 3s5p) + $Cs^+$ are significant and yield a pure linear behaviour function of R, especially at intermediate and large distance. For the remaining states $2^1\Sigma^+$, $4^1\Sigma^+$ , 6 $^1\Sigma^+$ and 10 $^1\Sigma^+$, dissociating into $Mg^+$ (3s, 3p, 4s, 3d, 4p, 5s, 4d, 4f, 5p, and 6s) + Cs(*6s, 6p, 4d, 5d,7s and 7p*), we note significant permanent dipole moments in specific regions and then they vanish swiftly at large distances.

For the permanent dipole moments of the first ten states of $^3\Sigma^+$ symmetry, similar observations were noticed as for the $^1\Sigma^+$ states. They exhibit undulations with abrupt variations at short and intermediate internuclear distance R such as between the two states $8^3\Sigma^+$ and $9^3\Sigma$ around 27.7 $a_0$. We remark that the permanent dipole moments of these states, one after other, behaves a straight line on the same curve and then drops to zero at particular distances corresponding to the avoided crossings between neighbour electronic states. In figure 10–12 we present, respectively, the permanent dipole moments of $^{1,3}\Pi$ and $^{1,3}\Delta$ symmetries. As already mentioned in the introduction, to our best knowledge no previous studies on dipole moments exist for the $(MgCs)^+$ cationic molecule. For that reason, we compare our results with similar alkali-alkaline earth ionic systems already published such as $(MgK)^+$ [56]and $(MgLi)^+$ [33]. As expected, we observed similar behaviour for the permanent dipole moments.

### III. 3. 2 Transition Dipole Moments

In addition, we have determined the transition dipole moment functions for the $(MgCs)^+$ionic molecular system between states of similar symmetries. As the transition dipole moments between adjacent states are the most significant, we present here only the transition between neighbours' states. They are displayed in Figures 13-15.

Theoretically, Fedorov et al [31] reported dipole moments for the molecular ions $(MgLi)^+$ and $(MgNa)^+$. They have calculated the transition dipole moments for the latter systems; however, no data for the $(MgCs)^+$ ionic molecule are available and these important physical quantities are presented here for the first time. We note a meaningful variation with a numerous extremum, which can be assigned to avoided crossing positions between adjacent states. We note also a change in the sign of some dipole transitions such as between $3^1\Sigma^+$-$4^1\Sigma^+$ states. This change in sign corresponds to sudden change in electronic wavefunctions, which is usually related to the avoided crossings between the potential energy curves.

For the triplet sigma states, the transition dipole moment curves between the neighbours' states are displayed in Figure 13. We note similar behaviours observed previously for the singlet states. For both multiplicities, $^{1,3}\Sigma$ states, the weakly avoided crossing series already found between neighbouring states lead to abrupt variations in the transition dipole moments.

For the $^{1,3}\Pi$ and $^{1,3}\Delta$ symmetries, the transition dipole moments are presented in figures 14-15. We note also the existence of peaks and abrupt variations for all these curves. These peaks are located at particular distances due to the avoided crossings in the adiabatic representation.

The analysis of these curves shows that, the variations of the transition dipole moments are important for short distances. At larger distances, the variations of the transition dipole moment between adiabatic states are weak and tend to zero or constant values, related to pure atomic transitions.

### III.4. Radiative lifetimes

In this section, we present the vibrational state lifetimes for $A^1\Sigma^+(2^1\Sigma^+)$ excited state. Given a vibrational level of the first excited electronic state two possible transitions can occur, bound-bound and bound-free transitions. The radiative lifetime of a vibrational level corresponding to only bound-bound transitions is given by:

$$\tau_{\upsilon'} = \frac{1}{\Gamma_{\upsilon'}} \text{ where } \quad \Gamma_{\upsilon'} = \sum_{\upsilon=0}^{nvt} A_{\upsilon\upsilon'} \quad (1)$$

$A_{\nu\nu'}$ is the Einstein coefficient linking for example the $A^1\Sigma^+(\nu')$ and $X^1\Sigma^+(\nu)$ levels. Zemke et al. [77] have showed, previously, that there is a missing contribution in the radiative lifetime. It corresponds to the bound-free transitions, which are more significant for the higher vibrational levels close to the dissociation limit of the excited electronic state, found in the form of the continuum radiation to states above the dissociation limit of the ground state. It corresponds to the contribution of the bound-free transition missed in the **Eq. (1).** It is related to the transition between the vibrational level $v'$, which belongs to the excited electronic state, to the continuum of the ground state or the lower state in general. Such contribution is not negligible as it was demonstrated by Partridge et al [76], Zemke et al [77] and Berriche and Gadea [78-79]. It is more important for the higher vibrational levels due the difference of the location of the potential wells. We calculate this contribution using two different approximations "Franck-Condon" [77-78]and "sum rule method" [79-80] approximation. In the Franck-Condon (FC) approximation proposed by Zemke et al [77], the bound-free contribution is given by:

$$A_{\upsilon'}(bound - free) = 2.14198639\,x10^{10}\left|\mu(R_{\upsilon'+})\right|^2 FC_{\upsilon',cont}(\Delta E)^3{}_{\upsilon',cont} \quad (2)$$

where $\Delta E_{\upsilon',cont} = E_{\upsilon'} - E_{as}$ is the energy difference between the vibrational level $v$and the energy of the asymptotic limit of the lower electronic state to which belongs the continuum.

The quantity $\mu(R_{\upsilon'+})$ corresponds to the transition dipole moment at the right external turning point of the vibrational level $v'$.

$$FC_{\upsilon',cont} = \int |\langle \chi_{\upsilon'}|\chi_E\rangle|^2 dE = 1 - \sum_{\upsilon=0}^{nvt} |\langle \chi_{\upsilon'}|\chi_\upsilon\rangle|^2 \quad (3)$$

The approximate sumrule approach, implemented by Pazyuk et al [81-82], allows producing the radiative lifetime components for diatomic vibronic states including the bound-free contribution. In addition, this approximation has a high efficiency for non-diagonal systems and particularly for those with significant continuum contributions. The radiative lifetime using the sum rule method is given by:

$$\frac{1}{\tau} = \int_{R_{\min}}^{R_{\max}} \varphi_{v'}(R)\left(D\left(R^2\right)\right)\left(\Delta\left(R^3\right)\right)\varphi_{v'}(R)dR \qquad (4)$$

$\varphi_{\upsilon'}(R)$ is the wavefunction of the vibrational level belonging to the $A^1\Sigma^+$ excited electronic state. D(R) is the transition dipole moment between the ground X $^1\Sigma^+$ and first excited $A^1\Sigma^+$states. ΔU(R) is the energy difference between the ground $X^1\Sigma^+$ and first excited $A^1\Sigma^+$ states. The total radiative lifetime, taking into account the bound-bound and bound-free contributions, using the two approaches, FC approximation and the approximate sum approach, are presented in **Table 4**.

To our best knowledge, the radiative lifetimes of the vibrational levels of the $A^1\Sigma^+$ state of the $(MgCs)^+$molecules are presented here for the first time. These radiative lifetimes are increasing with the vibrational level $v$ in a range of 10-43 ns.

### III.5. Ion-atom elastic collisions

In this section we study cold elastic collisions between an alkaline-earth atom and an alkali ion in the ground electronic state ($X^1\Sigma^+$) of $(MgCs)^+$ molecular system. The ground state potential asymptotically correlates to the heavier alkali element in the ionic state ($Cs^+$) and the alkaline element in its neutral ground state (Mg). The first singlet excited state $2^1\Sigma^+$ asymptotically corresponds to the charge-exchanged state of the $X^1\Sigma^+$ potential. We use Mg+ $Cs^+$ as a prototype system, several results equally apply to other alkaline-earth-alkali atom-ion collisional systems. The elastic collisions between the alkaline-earth atom, Mg, and alkali ion, $Cs^+$, for the ground state potential ($X^1\Sigma^+$) can be represented as:

$$Mg + Cs^+ \rightarrow \{MgCs^+(\mathrm{X}\Sigma^+)\} \rightarrow Mg + Cs^+ \qquad (5)$$

The partial-wave Schrödinger equation for the ion-atom elastic collision can be expressed as:

$$\left[\frac{d^2}{dr^2} + k^2 - \frac{2\mu}{\hbar^2}V(r) - \frac{l(l+1)}{r^2}\right]\psi_l(kr) = 0 \qquad (6)$$

where $r$ is the internuclear separation between atom and ion, $\mu$ is the reduced mass of an ion-atom colliding pair and $\psi_l(kr)$ is the wave function for the $l^{th}$ partial wave. The wave function satisfies the asymptotic boundary condition: $\psi_l(kr) \sim sin\left[kr - \frac{l\pi}{2} + \eta_l\right]$. The above second order differential Eq.6 can be solved by Numerov-Cooley algorithm [80] using three points recursion relation. Initially we set the values of the wave function at two initial consecutive points at small internuclear separation $r{\sim}0$. Thereafter, the code calculates the wave function at 3rd point by utilising the three point recursion relation. Then 2nd and 3rd points are reset to initial point and the wave function is calculated for the 4th point. This process is continued until the asymptotic boundary is attained. The initial boundary conditions are included by expanding the interaction potential at small $r$ and solved the Schrödinger equation at this limit. The boundary condition at asymptotic limit is incorporated by setting the condition that the effective long-range part of potential *V(r)* is much less than the collisional energy E (at least one tenth of E). We considered a constant step size 0.01 during the propagation of the code.

The total elastic scattering cross section is given by:

$$\sigma_{tot} = \frac{4\pi}{k^2}\sum_{l=0}^{\infty}(2l+1)sin^2(\eta_l) \qquad (7)$$

where $k$ is the wave number related with the collisional energy $E = \frac{\hbar^2 k^2}{2\mu}$ and $\eta_l$ being the phase shift for the$l$-th partial wave. As with increase in energy, a large number of partial wavesstart to contribute to elastic scattering cross section. Under this condition, the scattering cross section is approximated as[83]

$$\sigma_{tot} \approx \pi\left(\frac{\mu C_4^2}{\hbar^2}\right)^{1/3}\left(1+\frac{\pi^2}{16}\right)E^{-1/3} \quad (8)$$

But, at very low energy limit i.e$E \rightarrow 0$, the phase shift follows the Wigner threshold laws. At this energy regime, the phase shift is related to the $s$-wave scattering length $a_s = -\lim_{k\rightarrow 0}\frac{tan\eta_0}{k}$. The scattering length carries the information about the interaction potential of a system. A positive (negative) scattering length is associated with repulsive (attractive) interaction.

In figure 16, we show logarithm of total elastic scattering cross sections ($\sigma_{tot}$) in a.u. as a function of logarithm of energy $E$ in K for ground state collisions between neutral alkaline-earth atom Mg and alkali ion $Cs^+$. In order to calculate the convergence result sin$\sigma_{tot}$, we need 59 partial waves. In our calculations, we have included 70 partial waves. Since with the increase in energy, large number of partial waves start to contribute to total elastic scattering cross section. Considering logarithm on both sides of equation (8), a straight is obtained with slope equal to -1/3. From the linear fitting of log $\sigma_{tot}$ vs log E of figure 16, we have verified the numerically calculated slope of the straight line is quite close to the theoretical value.

**Sec.III.6: Two-photon Photoassociation: Molecular ion formation**

In this section, we explore the molecular ion formation in the ground state electronic potential $X^1\Sigma^+$ by two-photon incoherent Raman process in the presence of two applied lasers $L_1$ and $L_2$, respectively. The formation of molecular ion needs two steps. Initially the atom-ion pair present in the ground state scattering continuum ($X^1\Sigma^+$) is irradiated with laser $L_1$ with a suitable frequency to form the molecular ion in one of the bound levels of first excited potential $2^1\Sigma^+$. Once the molecular ion is formed, then laser $L_2$ is turned on to de-excite the molecular ion to the ground electronic potential followed by bound-bound transition.

Here, the rate coefficient of the first concerned step, which is known as single-photon Photoassociation (PA) at a temperature T, is given by [84]

$$K_{PA} = \left\langle \frac{\pi v_{rel}}{k^2} \sum_{l=0}^{\infty} (2l+1) |S_{PA}(E, l, \omega_l)|^2 \right\rangle \quad (9)$$

where $v_{\mathrm{rel}}$ is the relative velocity of the interacting particles, $S_{PA}$ is the scattering matrix element, and $< \cdots >$implies an averaging over thermal velocity distribution. At very low temperature limit, the energy of the interacting particle is such that their dynamics may be described by the *s*-wave scattering. If one considers a Maxwellian velocity distribution of an ultra-cold mixture of gas, the equation (9) can be represented as:

$$K_{PA}(T, \omega_L) = \frac{1}{hQ_T} \sum_{l=0}^{\infty} (2l+1) \int_0^{\infty} |S_{PA}(E, l, \omega_L)|^2 e^{-\beta E} dE \quad (10)$$

Here, $Q_T = \left(2\pi\mu k_B T / h^2\right)^{3/2}$ is the translational partition function, $\beta = 1/k_b T$, and $k_B$ is the Boltzmann constant. The scattering *S* matrix element is given by

$$|S_{PA}|^2 = \frac{\gamma_s \Gamma_l}{\delta_E^2 + (\gamma/2)^2} \quad (11)$$

The scattering phase shift (η) is related with scattering T-matrix as $T = \frac{\mathbf{1} - e^{2i\eta}}{2i} = -e^{i\eta} \sin\eta$ while the quantity $|T|^2 = \sin^2\eta$. Again, the scattering S matrix is related with the T matrix as $S = \mathbf{1} - 2iT$ and the quantity $|S|^2 = \mathbf{1} + 4\sin^2\eta$. Therefore, in term of phase shift and T-matrix, the Eq.11 becomes

$$|T|^2 = \sin^2\eta = \frac{1}{4}\left[\frac{\gamma_s \Gamma_l}{\delta_E^2 + (\gamma/2)^2} - \mathbf{1}\right] \quad (12)$$

where detuning parameters are defined as $\delta_E = \frac{E}{\hbar} + \delta_{v,j}$, $\delta_{v,j} = \omega_L - \omega_{v,j}$ with $E_{v,j} = \hbar\omega_{v,j}$ is the binding energy of the excited ro-vibrational state, $\omega_L$ is the frequency of the applied laser $L_1$. Here, $\gamma_s$ is the spontaneous line width and $\Gamma_l$ is the stimulated line width. We consider that other decay processes, such as molecular predissociation are either negligible or not present in our system.

The PA rate is determined, basically, by stimulated line width and can be expressed by the golden rule as

$$\Gamma_l = \frac{\pi I}{\varepsilon_0 c} \left| \left\langle \phi_{v,j} | D(r) | \psi_l(kr) \right\rangle \right|^2 \quad (13)$$

We termed the quantity $D_{v,j,l} = \langle \phi_{v,j} | D(r) | \psi_l(kr) \rangle$ as the radial transition dipole matrix element between the unit normalized bound state wave function $\phi_{v,j}$ and energy normalized continuum wave function $\psi_l(kr)$, and $D(r)$ is the transition dipole moment. Here, $I$ is the intensity of the laser$L_1$, $c$ is the velocity of light, and $\varepsilon_0$is the vacuum permittivity.

In order to calculate the quantity $D_{v,j,l}$, for optimal production of molecular ion by PA process, we have to calculate the scattering state and suitable bound state wave functions. We solve the partial wave Schrödinger equation (6) by standard renormalized Numerov-Cooley method [80]for determination of the wavefunctions that we discussed. In general, molecular dipole transitions between two bound states or between continuum and bound states is controlled by the Franck-Condon (FC) principle. The free-bound FC value is highest for the bound state $\phi_{v,j}(v=17, j=1)$ of the first excited potential $2^1\Sigma^+$of $(MgCs)^+$ system. The PA rate coefficient is calculated according to equation (9) to form the molecular ion in the excited electronic state $2^1\Sigma^+$. In figure 17, we have shown the rate coefficient of PA as a function of energy $E$ (in Kelvin) considering the atom-ion pairs Mg-$Cs^+$ are initially lying in the ground state scattering continuum. We considered the laser intensity $L_1$ = 10 W/cm$^2$ tuned at the PA resonance. We find that our calculated PA rate coefficient is consistent with the results of $Li^+$-Be [83] and $K^+$-Mg [56] systems. Finally, in figure 18, we have shown the variation of the rate of ion-atom PA as a function of intensity of the applied laser$L_1$ considering the temperature at 10μK. We have observed that a saturation effect on ion-atom PA rate occurs near the laser intensity I = 35 kW/cm$^2$.

Now we discuss about the formation of a ground state molecular ion by stimulated Raman type process in the presence of a second applied laser $L_2$ tuned near a bound-bound transition between the potentials $2^1\Sigma^+$and $X^1\Sigma^+$, respectively. To find the efficacy of the coherent laser coupling between the two bound states, we calculate the Rabi coupling Ω given by

$$\hbar\Omega = \left(\frac{I}{4\pi c\varepsilon_0}\right)^{1/2} \left|\langle \phi_{v,j} | \boldsymbol{D}(r).\boldsymbol{\epsilon} | \phi_{v'j'} \rangle\right| (14)$$

where $I$ is the intensity of the second applied laser$L_2$ tuned to the transition frequency, $\boldsymbol{\epsilon}$is the unit vector of laser polarisation, and $\phi_{v,j}$ and $\phi_{v'j'}$ are the two bound states of excited ($2^1\Sigma^+$) and ground ($X^1\Sigma^+$) state electronic potentials, respectively. We have calculated the Rabi frequencies between the selected bound state ($v=15$, $j=1$) of the first excited molecular potential and different bound states of the ground electronic potential considering an intensity of laser $L_2$ = 10 kW/cm$^{-2}$. We observed that the bound-bound Rabi coupling is maximum (Ω =

150 MHz) for the bound state $\phi_{v'j'}(v'=16,\ j'=0)$ of $X^1\Sigma^+$ potential. Comparing this value with spontaneous line width ($\gamma$ = 120 kHz) of excited bound state, calculated using the formula [83], we found that bound-bound Rabi coupling ($\Omega$) is several orders of magnitude higher. Therefore, we may infer that the ground state molecular ion formation is likely to be possible by stimulated Raman type process [56,85,86].

## Conclusion

In the present work, an extended and complete study devoted to the $(MgCs)^+$ molecular ion was presented. We have performed precise ab initio calculations for the potential energy curves and their related spectroscopic constants for the ground and the 41 low-lying excited states of $^{1,3}\Sigma^+$, $^{1,3}\Pi$ and $^{1,3}\Delta$ symmetries below the $Mg^{2+}Cs^-$ asymptotic limit.

The used calculation method is based on a non-empirical pseudo-potential approach for $Mg^{2+}$ and $Cs^+$ cores, which allows treating the ionic molecule as a 2-electron system where full configuration interaction FCI calculations can be easily performed. Interesting series of avoided crossings have been observed between same symmetry of the $^{1,3}\Sigma^+$, $^{1,3}\Pi$ states. These avoided crossings often lead to relatively irregular dipole moment curves, which are both manifestations of abrupt changes of the character of the electronic wave functions with underlying strong interactions between the electronic states. These interactions could give rise to important charge transfer for collisions between the atom–ion combinations. In addition, the vibrational levels and their spacings were investigated for all attractive states. Finally, electric permanent and transition dipole moments were calculated. As it is expected the permanent dipole moments of the electronic states dissociating into Mg {(3s$^2$ ($^1$S), 3s3p ($^3$P), 3s3p ($^1$P), 3s4s ($^3$S), 3s4s ($^1$S), 3s3d ($^1$D), 3s4p ($^3$P), 3s3d ($^3$D), ($^1$P), 3s5s ($^3$S), 3s5s ($^1$S), 3s4d ($^1$D), 3s4d ($^3$D), and 3s5p ($^3$P))}+$Cs^+$ have shown an almost linear features function of R, especially for intermediate and large internuclear distances. Moreover, the abrupt changes in the permanent dipole moment are localized at particular distances corresponding to the avoided crossings between the two neighbour electronic states.

In this study, the ab initio calculations were carefully performed to produce accurate data, though the lack of experimental studies for this molecular ion. However, theoretically only the ground state of $(MgCs)^+$ was studied in a recent work published by Smialkowski and Tomza [32]. We expect that our data can serve as a solid foundation for future and advanced dynamic studies for $(MgCs)^+$ molecular ion.

Then, these precise data were used for elastic scattering calculations between Mg and $Cs^+$ at low temperatures. In fact, theoretical understanding of low-energy atom-ion scattering may help for probing dynamics of quantum gases. Besides that, we predict theoretically the formation of a translationally and rotationally cold molecular ion by two-photon incoherent Raman process. The formation of such cold molecular ions is very important in the research domain of ultra-cold chemistry and precision spectroscopy. We anticipate that our analysis including the ab initio data may serve as a solid platform for future study on $(MgCs)^+$ molecular ionic system.

**Tables and Figures captions**

**Table 1**: Polarizabilities, cut-off radii and pseudo-potential parameters (in a.u.) of Mg and Cs atoms.

**Table 2:** Asymptotic energies of the alkali earth $(MgCs)^+$ electronic states (in a.u.): comparison between our calculated energy, Moore et al [73] energies and the corresponding experimental dissociation

**Table 3:** Spectroscopic constants of the ground and excited singlet and triplet electronic states of $(MgCs)^+$ molecule

**Table 4:** Radiative lifetimes and vibrational-level spacing of the 2 $^1\Sigma^+$state of the $(MgCs)^+$ ion.

**Figure 1:** Adiabatic potential energy curves of the first $10^1\Sigma^+$ electronic states of $(MgCs)^+$

**Figure 2:** Adiabatic potential energy curves of the first $10^3\Sigma^+$ electronic states of $(MgCs)^+$

**Figure 3:** Adiabatic potential energy curves of the lowest $^1\Pi$ of the alkali earth $(MgCs)^+$

**Figure 4:** Adiabatic potential energy curves of the lowest $^3\Pi$ of the alkali earth $(MgCs)^+$

**Figure 5:** Adiabatic potential energy curves of the three lowest $^1\Delta$ (solid line) states and three lowest $^3\Delta$ (dashed line) states of the alkali earth $(MgCs)^+$

**Figure 6:** Vibrational energy level spacing ($E(v+1)–E(v)$) (in $cm^{-1}$) for the ($^{1,3}\Sigma^+$) electronic states of the $(MgCs)^+$molecule.

**Figure 7:** Vibrational energy level spacing ($E(v+1)–E(v)$) (in $cm^{-1}$) for the $^{1,3}\Pi$ electronic states of the $(MgCs)^+$molecule.

Figure 8. (a) Permanent dipole moment for the $^1\Sigma^+$ states of the $(MgCs)^+$ molecular ion

(b) Zoomed area at distances between 0 and 15 a.u.

Figure 9. (a) Permanent dipole moment for the $^3\Sigma^+$ states of the $(MgCs)^+$ molecular ion.

(b) Zoomed area at distances between 0 and 15 a.u.

Figure 10. (a) Permanent dipole moment for the $^1\Pi$ states of the $(MgCs)^+$ molecular ion.

(b) Zoomed area at distances between 0 and 15 a.u.

Figure 11. (a) Permanent dipole moment for the $^3\Pi$ states of the $(MgCs)^+$ molecular ion.

(b) Zoomed area at distances between 0 and 15 a.u.

**Figure 12:** Permanent dipole moment for the $^{1,3}\Delta$ states of the $(MgCs)^+$ molecular ion as a function of the inter-nuclear distance $R$.

**Figure 13:** Transition dipole moments for selected (1-2, 2-3 and 3-4) $^{1,3}\Sigma^{+}$ states, as a function of the internuclear distance *R*.

**Figure 14:** Transition dipole moments for selected (1-2, 2-3 and 3-4) $^{1,3}\Pi$ states, as a function of the internuclear distance *R*.

**Figure 15:** Transition dipole moments for selected (1-2, 1-3and 2-3) $^{1,3}\Delta$states, as a function of the internuclear distance *R*.

**Figure 16:** The variation of T-matrix element as a function of collisional energy E in Kelvin.

**Figure 17:** The rate constant $K_{PA}$ ($cm^3s^{-1}$) of PA as a function of energy *E* (in Kelvin) for laser frequency 10 $W/cm^2$ tuned at PA resonance.

**Figure 18:** The variation of the rate constant of Photoassociation as function of the intensity of the applied laser.

**Table 1**: Polarizabilities, cut-off radii and pseudo-potential parameters (in a.u.) atoms.

| Atom | α | $\rho_s$ | $\rho_p$ | $\rho_d$ |
|---|---|---|---|---|
| Mg(Z=12) | 0.46904 | 0.9 | 1.25499 | 1.500 |
| Cs(Z=55) | 15.117 | 2.69 | 1.85 | 2.810 |

**Table 2:** Asymptotic energy of the alkali earth $(MgCs)^+$ electronic states (in $cm^{-1}$): comparison between our calculated energy, Moore et al [73] calculated energy and the corresponding experimental dissociation

| State | Asymptotic molecular state | This work | Experiment [73] |
|---|---|---|---|
| X (1) $^1\Sigma^+$ | $Mg(3s^2)+Cs^+$ | 0 | 0 |
| A (2) $^1\Sigma^+$ | $Mg^+(3s)+Cs(6s)$ | 30263.62 | 30398.60 |
| C (3) $^1\Sigma^+$ | $Mg(3s3p)+Cs^+$ | 35050.59 | 35319.67 |
| D (4) $^1\Sigma^+$ | $Mg^+(3s)+Cs(6p)$ | 41811.30 | 41946.72 |
| E (5) $^1\Sigma^+$ | $Mg(3s4s)+Cs^+$ | 43502.57 | 43472.29 |
| F (6) $^1\Sigma^+$ | $Mg^+(3s)+Cs(5d)$ | 44821.40 | 44956.60 |
| G (7) $^1\Sigma^+$ | $Mg(3s3d)+Cs^+$ | 46402.50 | 47967.79 |
| H (8) $^1\Sigma^+$ | $Mg^+(3s)+Cs(7s)$ | 48802.02 | 48934.58 |
| I (9) $^1\Sigma^+$ | $Mg(3s4p)+Cs^+$ | 49346.10 | 49487.00 |
| J (10) $^1\Sigma^+$ | $Mg^+(3s)+Cs(7p)$ | 52143.74 | 52284.65 |
| a (1) $^3\Sigma^+$ | $Mg(3s3p)+Cs^+$ | 21876.60 | 21994.46 |
| c (2) $^3\Sigma^+$ | $Mg^+(3s)+Cs(6s)$ | 30263.62 | 30398.60 |
| d (3) $^3\Sigma^+$ | $Mg(3s4s)+Cs^+$ | 41196.77 | 41154.85 |
| e (4) $^3\Sigma^+$ | $Mg^+(3s)+Cs(6p)$ | 41811.30 | 41946.72 |
| f (5) $^3\Sigma^+$ | $Mg^+(3s)+Cs(5d)$ | 44821.40 | 44956.60 |
| g (6) $^3\Sigma^+$ | $Mg(3s4p)+Cs^+$ | 47857.84 | 48197.80 |
| h (7) $^3\Sigma^+$ | $Mg(3s3d)+Cs^+$ | 47956.38 | 46647.87 |
| I (8) $^3\Sigma^+$ | $Mg^+(3s)+Cs(7s)$ | 48802.02 | 48934.58 |
| J (9) $^3\Sigma^+$ | $Mg(3s5s)+Cs^+$ | 51871.81 | 51976.72 |
| k (10) $^3\Sigma^+$ | $Mg^+(3s)+Cs(7p)$ | 52143.74 | 52284.65 |
| $B^1\Pi$ | $Mg(3s3p)+Cs^+$ | 35050.59 | 35319.67 |
| $2^1\Pi$ | $Mg^+(3s)+Cs(6p)$ | 41811.30 | 41946.72 |
| $3^1\Pi$ | $Mg^+(3s)+Cs(5d)$ | 44821.40 | 44956.60 |
| $4^1\Pi$ | $Mg(3s3d)+Cs^+$ | 46402.50 | 47967.79 |
| $5^1\Pi$ | $Mg(3s4p)+Cs^+$ | 49346.10 | 49487.00 |
| $6^1\Pi$ | $Mg^+(3s)+Cs(7p)$ | 52143.74 | 52284.65 |
| $7^1\Pi$ | $Mg^+(3s)+Cs(6d)$ | 52948.78 | 53013.74 |
| $8^1\Pi$ | $Mg(3s4d)+Cs^+$ | 53134.01 | 54397.75 |
| $b^3\Pi$ | $Mg(3s3p)+Cs^+$ | 21876.60 | 21994.46 |
| $2^3\Pi$ | $Mg^+(3s)+Cs(6p)$ | 41811.30 | 41946.72 |
| $3^3\Pi$ | $Mg^+(3s)+Cs(5d)$ | 44821.40 | 44956.60 |
| $4^3\Pi$ | $Mg(3s4p)+Cs^+$ | 47857.84 | 48197.80 |
| $5^3\Pi$ | $Mg(3s3d)+Cs^+$ | 47956.38 | 46647.87 |
| $6^3\Pi$ | $Mg^+(3s)+Cs(7p)$ | 52143.74 | 52284.65 |
| $7^3\Pi$ | $Mg^+(3s)+Cs(6d)$ | 52948.78 | 53013.74 |
| $8^3\Pi$ | $Mg(3s5p)+Cs^+$ | 54249.83 | 54900.57 |
| $1^1\Delta$ | $Mg(3s3p)+Cs^+$ | 35050.59 | 35319.67 |
| $2^1\Delta$ | $Mg^+(3s)+Cs(5d)$ | 44821.40 | 44956.60 |
| $3^1\Delta$ | $Mg(3s3d)+Cs^+$ | 46402.50 | 47967.79 |
| $1^3\Delta$ | $Mg(3s3p)+Cs^+$ | 21876.60 | 21994.46 |
| $2^3\Delta$ | $Mg^+(3s)+Cs(5d)$ | 44821.40 | 44956.60 |
| $3^3\Delta$ | $Mg(3s4p)+Cs^+$ | 47857.84 | 48197.80 |

**Table 3:** Spectroscopic constants of the ground and excited singlet and triplet electronic states of $(MgCs)^+$ molecule

| State | $R_e$(a. u.) | $D_e$(cm$^{-1}$) | $T_e$(cm$^{-1}$) | $\omega_e$(cm$^{-1}$) | $\omega_e x_e$(cm$^{-1}$) | $B_e$(cm$^{-1}$) |
|---|---|---|---|---|---|---|
| $X^1\Sigma^+$ | 7.70 | 2047 | 0 | 73.2 | 0.63 | 0.049364 |
| | 7.85[30] | 1861[32] | | 73.2[32] | 0.0481[32] | |
| $2^1\Sigma^+$ | 11.45 | 2662 | 29782 | 44.64 | 0.25 | 0.022354 |
| $3^1\Sigma^+$ | 16.88 | 1054 | 36391 | 21.92 | 0.16 | 0.010280 |
| $4^1\Sigma^+$ | 9.33 | -2750 | | | | |
| | 19.46 | 1891 | 42101 | 21.73 | 0.06 | 0.007700 |
| $5^1\Sigma^+$ | 28.24 | 263 | 45256 | 9.12 | 0.07 | 0.003674 |
| $6^1\Sigma^+$ | 58.89 | 4 | 46990 | -0.48 | 14.61 | 0.001160 |
| $7^1\Sigma^+$ | 8.68 | -3301 | | | | |
| | 27.83 | 509 | 48507 | 8.92 | 26.51 | 0.038919 |
| $8^1\Sigma^+$ | 8.68 | -8647 | | | | |
| | 31.99 | 923 | 50059 | 10.13 | 29.95 | 0.038864 |
| $9^1\Sigma^+$ | 8.69 | -10460 | | | | |
| | 42.75 | 300 | 51240 | 5.44 | 30.84 | 0.038839 |
| $10^1\Sigma^+$ | 8.71 | -10297 | | | | |
| | 42.27 | 1105 | 53289 | 8.07 | 40.10 | 0.038614 |
| $1^1\Pi$ | 8.18 | 1087 | 36358 | 46.59 | 0.86 | 0.043818 |
| $2^1\Pi$ | 14.50 | 800 | 5886 | 21.99 | 1.39 | 0.013923 |
| $3^1\Pi$ | Repulsive | | | | | |
| $4^1\Pi$ | 22.97 | 356 48661 | 9.79 | 0.08 | 0.005555 | |
| $5^1\Pi$ | | Repulsive | | | | |
| $6^1\Pi$ | 33.43 | 613 | 53788 | 8.16 | 0.02 | 0.002622 |
| $7^1\Pi$ | 45.12 | -99 | | | | |
| $8^1\Pi$ | 59.83 | 66 | 55460 | 2.68 | 0.02 | 0.000819 |
| $1^1\Delta$ | Repulsive | | | | | |
| $2^1\Delta$ | Repulsive | | | | | |
| $3^1\Delta$ | Repulsive | | | | | |
| $1^3\Sigma^+$ | 8.58 | 4504 | 19556 | 74.78 | 0.52 | 0.039826 |
| $2^3\Sigma^+$ | 14.98 | 813 | 31630 | 25.56 | 0.17 | 0.013050 |
| $3^3\Sigma^+$ | 17.20 | 2658 | 40542 | 26.85 | 0.05 | 0.009905 |
| $4^3\Sigma^+$ | 28.15 | 506 | 43486 | 10.12 | 0.04 | 0.003698 |
| $5^3\Sigma^+$ | 58.89 | 4 | 46990 | -0.48 | 14.61 | 0.001160 |
| $6^3\Sigma^+$ | 28.13 | 1600 | 48413 | 13.19 | 0.02 | 0.003704 |
| $7^3\Sigma^+$ | 38.20 | 31 | 50667 | 6.81 | 0.03 | 0.002008 |
| $8^3\Sigma^+$ | 49.32 | 131 | 50850 | 4.59 | 0.02 | 0.001205 |
| $9^3\Sigma^+$ | 40.09 | 1055 | 52967 | 8.79 | 0.01 | 0.001823 |
| $10^3\Sigma^+$ | 56.76 | 395 | 53999 | 3.95 | 0.01 | 0.000910 |
| $1^3\Pi$ | 7.56 | 649 | 23411 | 55.73 | 1.49 | 0.051268 |
| $2^3\Pi$ | 13.19 | 1203 | 42791 | 26.38 | 0.46 | 0.016842 |
| $3^3\Pi$ | Repulsive | | | | | |
| $4^3\Pi$ | 23.35 | 772 | 49244 | 11.99 | 0.06 | 0.005376 |
| $5^3\Pi$ | 9.41 | -6463 | | | | |
| $6^3\Pi$ | 10.86 | -3328 | | | | |
| | 32.76 | 613 | 53788 | 8.23 | 16.29 | 0.024855 |
| $7^3\Pi$ | 14.23 | -4048 | | | | |
| $8^3\Pi$ | 16.49 | -3258 | | | | |
| | 45.30 | 250 | 56195 | 4.84 | 15.20 | 0.010767 |

| | |
|---|---|
| $1^3\Delta$ | Repulsive |
| $2^3\Delta$ | Repulsive |
| $3^3\Delta$ | Repulsive |

**Table 4.** Radiative lifetimes and vibrational-level spacing $E_v - E_{v-1}$ of the $2^1\Sigma^+$state of the $(MgCs)^+$ ion.

| Vibrational level | $E_v$-$E_{v-1}$($cm^{-1}$) | Frank Condon (ns) | Sumrule (ns) |
|---|---|---|---|
| 0 | | 10.579 | 10.582 |
| 1 | 44.110 | 10.658 | 10.661 |
| 2 | 43.878 | 10.739 | 10.742 |
| 3 | 43.599 | 10.825 | 10.828 |
| 4 | 43.367 | 10.912 | 10.915 |
| 5 | 43.080 | 11.003 | 11.006 |
| 6 | 42.848 | 11.096 | 11.098 |
| 7 | 42.550 | 11.191 | 11.194 |
| 8 | 42.278 | 11.290 | 11.294 |
| 9 | 42.016 | 11.394 | 11.399 |
| 10 | 41.744 | 11.502 | 11.511 |
| 11 | 41.477 | 11.634 | 11.666 |
| 12 | 41.213 | 11.770 | 11.824 |
| 13 | 40.941 | 12.055 | 12.253 |
| 14 | 40.666 | 12.294 | 12.589 |
| 15 | 40.389 | 12.989 | 13.787 |
| 16 | 40.109 | 13.549 | 14.649 |
| 17 | 39.832 | 14.486 | 16.450 |
| 18 | 39.534 | 15.301 | 17.708 |
| 19 | 39.237 | 15.689 | 18.283 |
| 20 | 38.926 | 15.967 | 18.465 |
| 21 | 38.604 | 16.259 | 18.689 |
| 22 | 38.293 | 17.042 | 19.857 |
| 23 | 37.992 | 17.532 | 20.454 |
| 24 | 37.668 | 18.321 | 21.644 |
| 25 | 37.352 | 18.660 | 21.795 |
| 26 | 37.041 | 19.142 | 22.251 |
| 27 | 36.707 | 19.728 | 22.916 |
| 28 | 36.381 | 20.504 | 23.969 |
| 29 | 36.044 | 21.158 | 24.555 |
| 30 | 35.699 | 21.574 | 24.792 |
| 31 | 35.361 | 22.446 | 25.859 |
| 32 | 35.012 | 23.075 | 26.408 |
| 33 | 34.666 | 23.816 | 27.190 |
| 34 | 34.312 | 24.277 | 27.467 |
| 35 | 33.959 | 25.217 | 28.468 |
| 36 | 33.599 | 26.400 | 29.687 |
| 37 | 33.235 | 26.976 | 30.140 |
| 38 | 32.868 | 27.668 | 30.647 |
| 39 | 32.493 | 28.386 | 31.252 |
| 40 | 32.114 | 29.689 | 32.700 |
| 41 | 31.726 | 30.554 | 33.375 |
| 42 | 31.333 | 31.446 | 34.124 |
| 43 | 30.932 | 32.939 | 35.617 |
| 44 | 30.526 | 33.864 | 36.390 |
| 45 | 30.118 | 34.807 | 37.156 |
| 46 | 29.705 | 35.796 | 37.999 |
| 47 | 29.287 | 37.175 | 39.363 |
| 48 | 28.863 | 38.664 | 40.724 |
| 49 | 28.435 | 39.788 | 41.692 |
| 50 | 28.000 | 41.625 | 43.490 |

**Figure 1 :**Adiabatic potential energy curves of the first $10^1\Sigma^+$ electronic states of $(MgCs)^+$

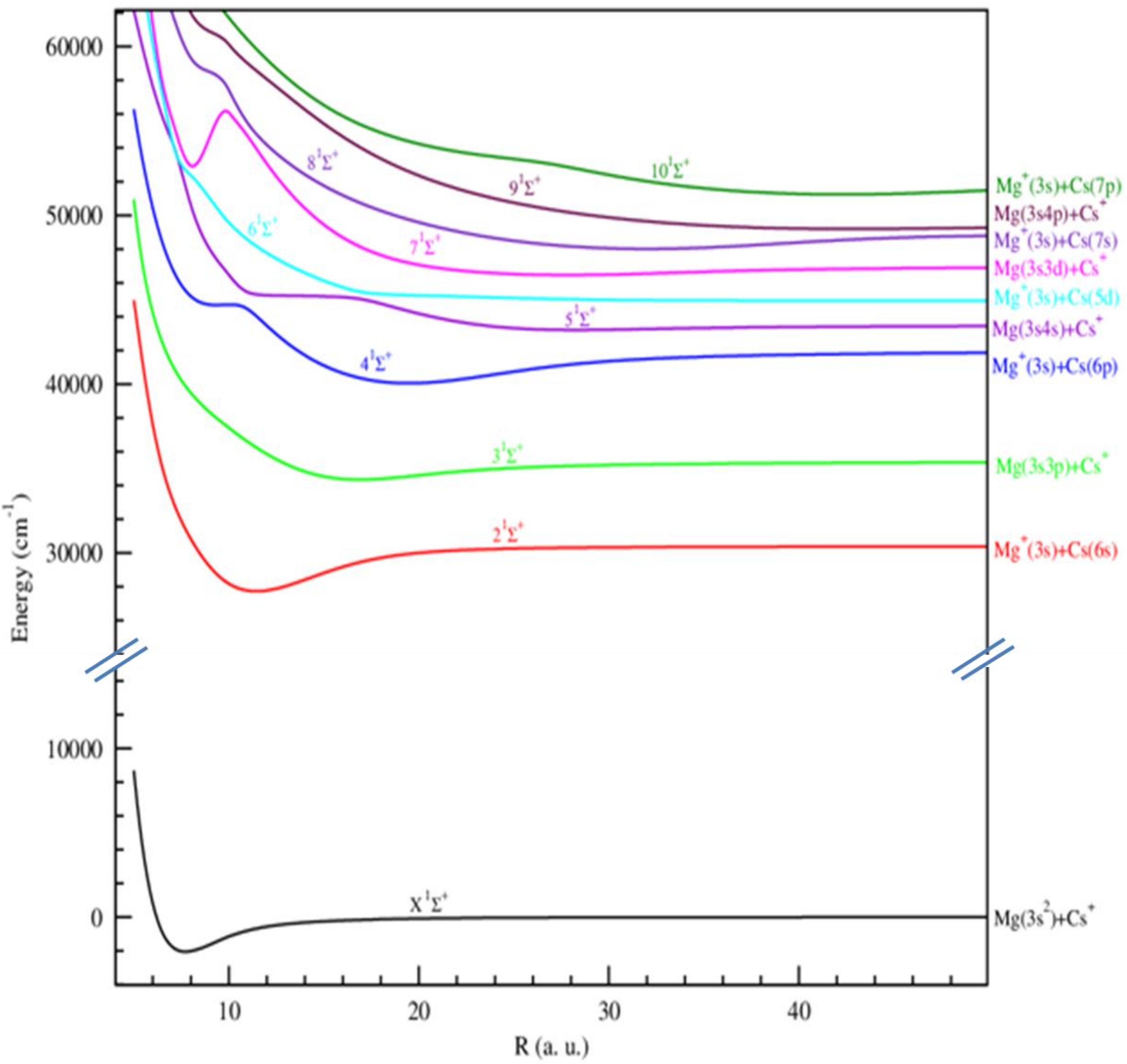

**Figure 2 :**Adiabatic potential energy curves of the first $10^3\Sigma^+$ electronic states of $(MgCs)^+$

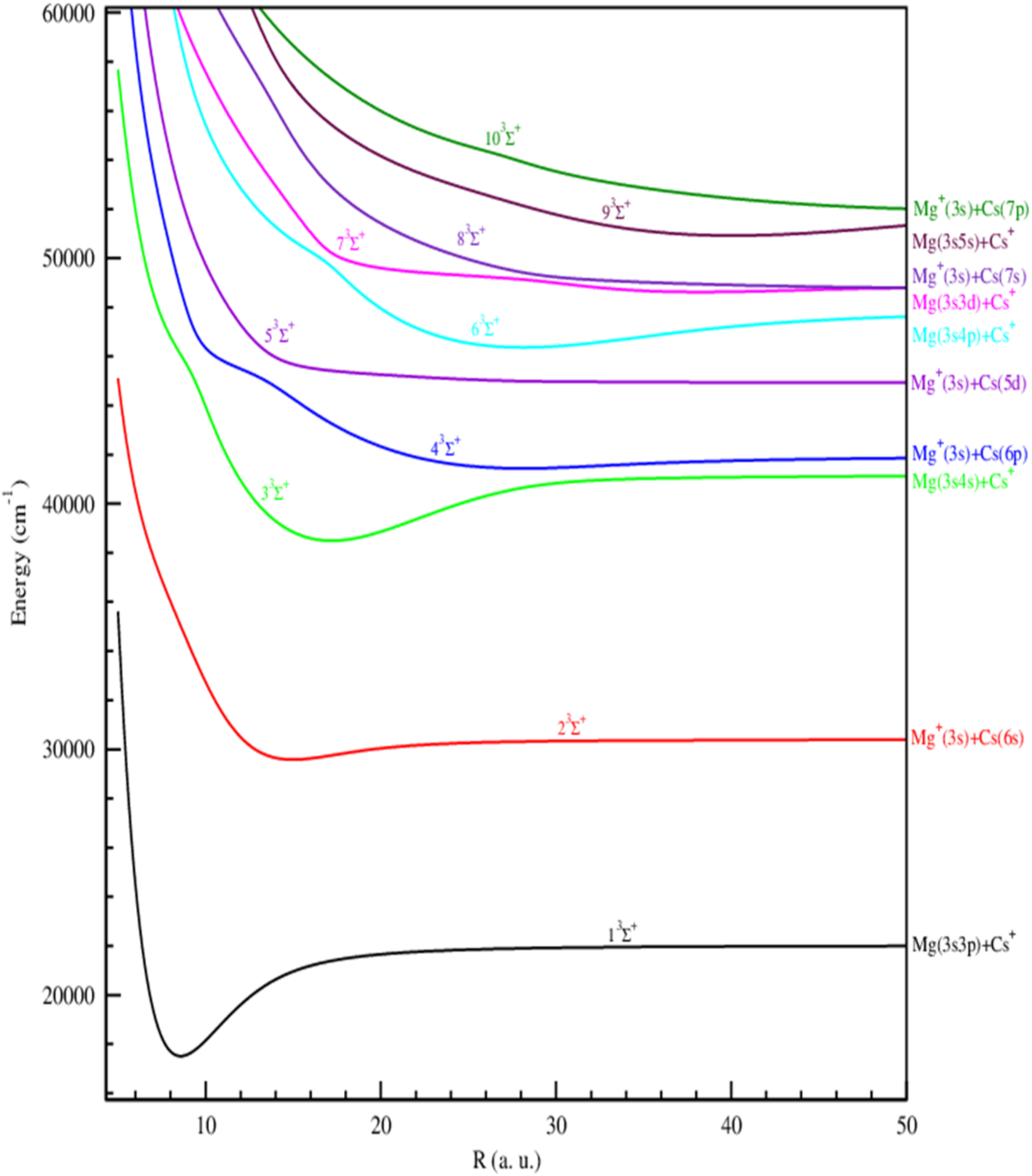

**Figure 3:** Adiabatic potential energy curves of the lowest $^1\Pi$of the alkali earth $(MgCs)^+$

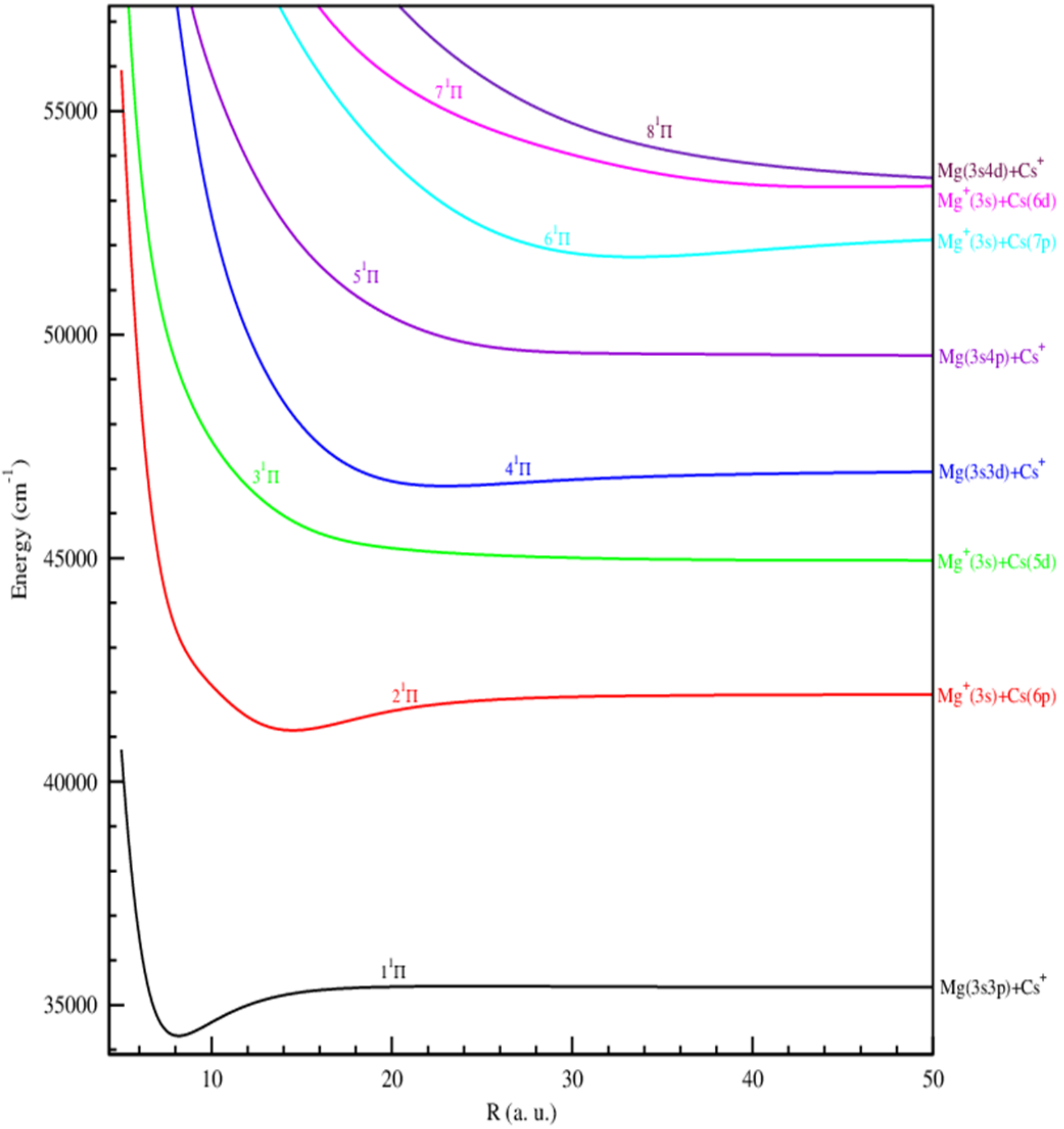

**Figure 4:** Adiabatic potential energy curves of the lowest $^3\Pi$of the alkali earth $(MgCs)^+$

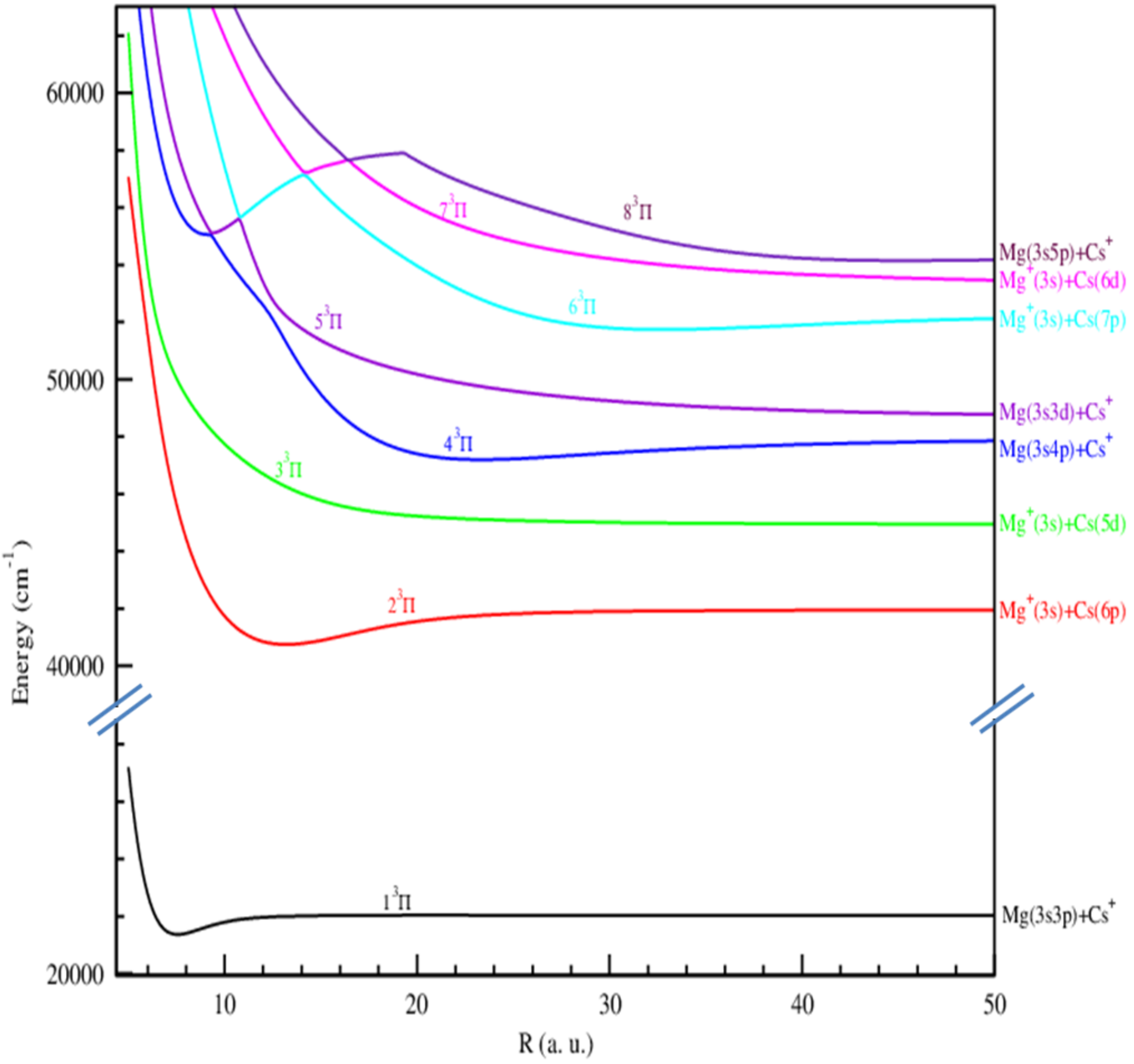

**Figure 5:** Adiabatic potential energy curves of the three lowest $^1\Delta$ (solid line) states and three lowest $^3\Delta$ (dashed line) states of the alkali earth $(MgCs)^+$

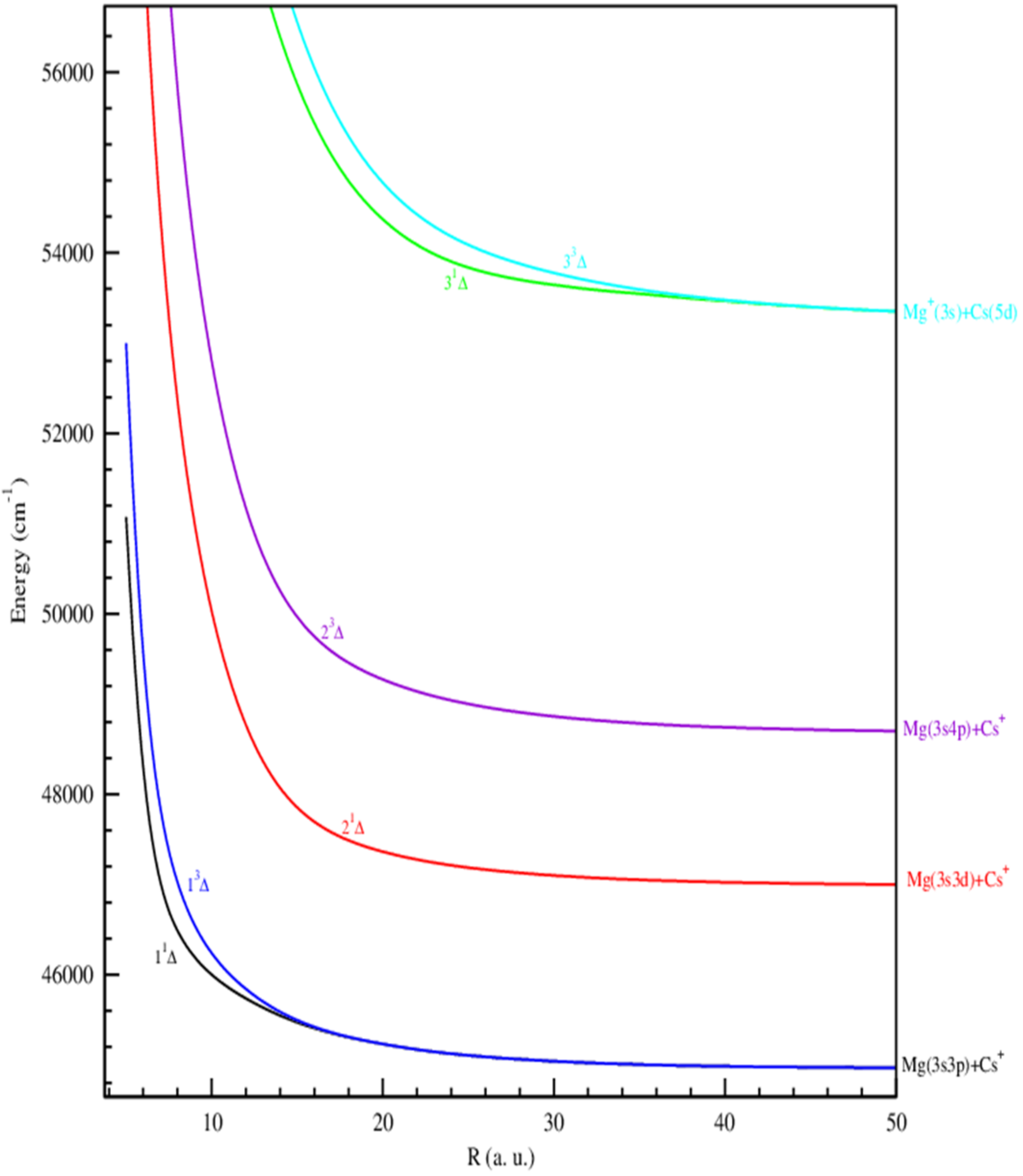

**Figure 6.** Vibrational energy level spacing ($E(\upsilon+1)$–$E(\upsilon)$) (in $cm^{-1}$) for the ($^{1,3}\Sigma^{+}$) electronic states of the $(MgCs)^{+}$molecule.

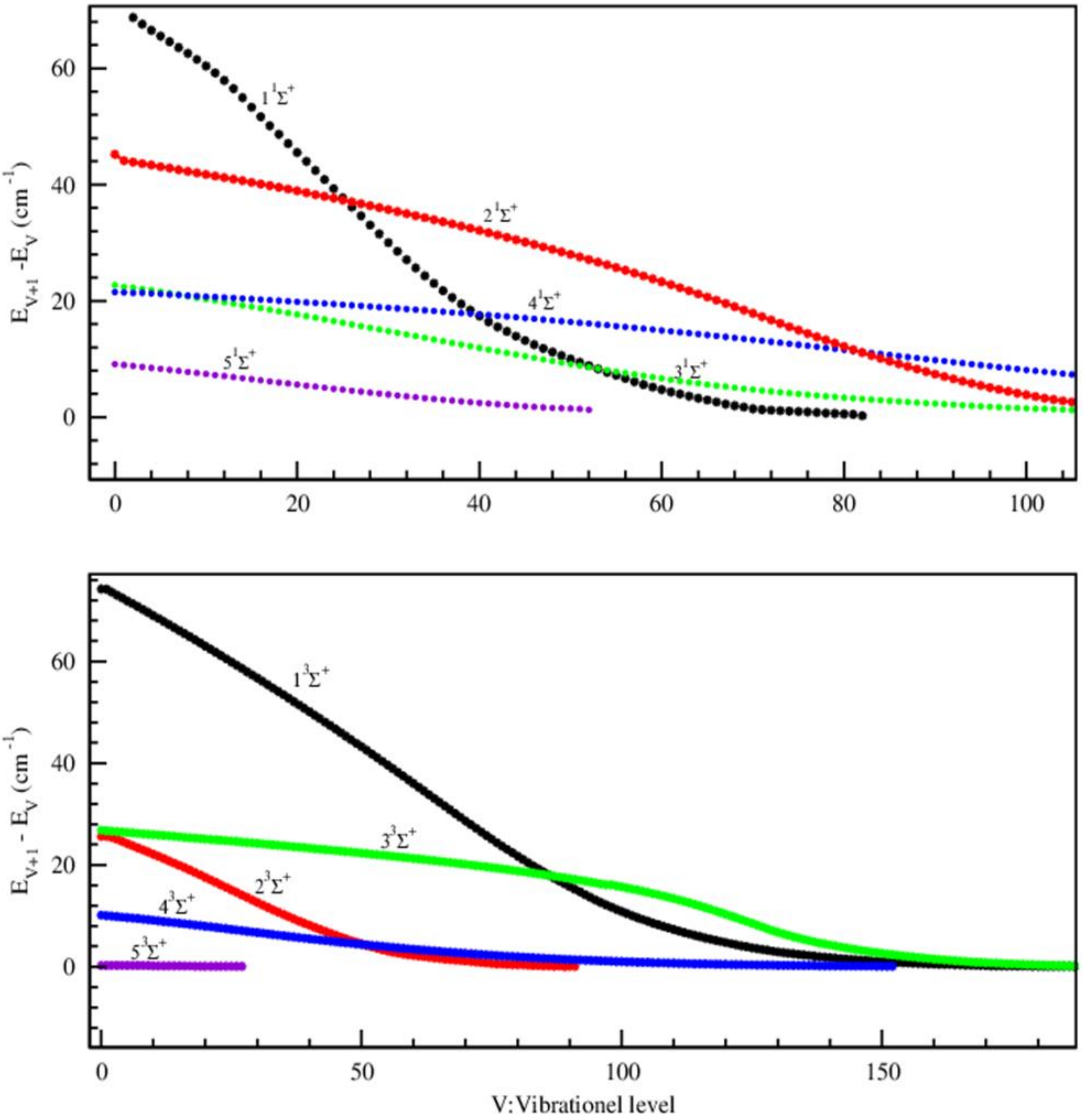

**Figure 7.** Vibrational energy level spacing ($E(\upsilon+1)$–$E(\upsilon)$) (in $cm^{-1}$) for the $^{1,3}\Pi$ electronic states of the $(MgCs)^+$molecule.

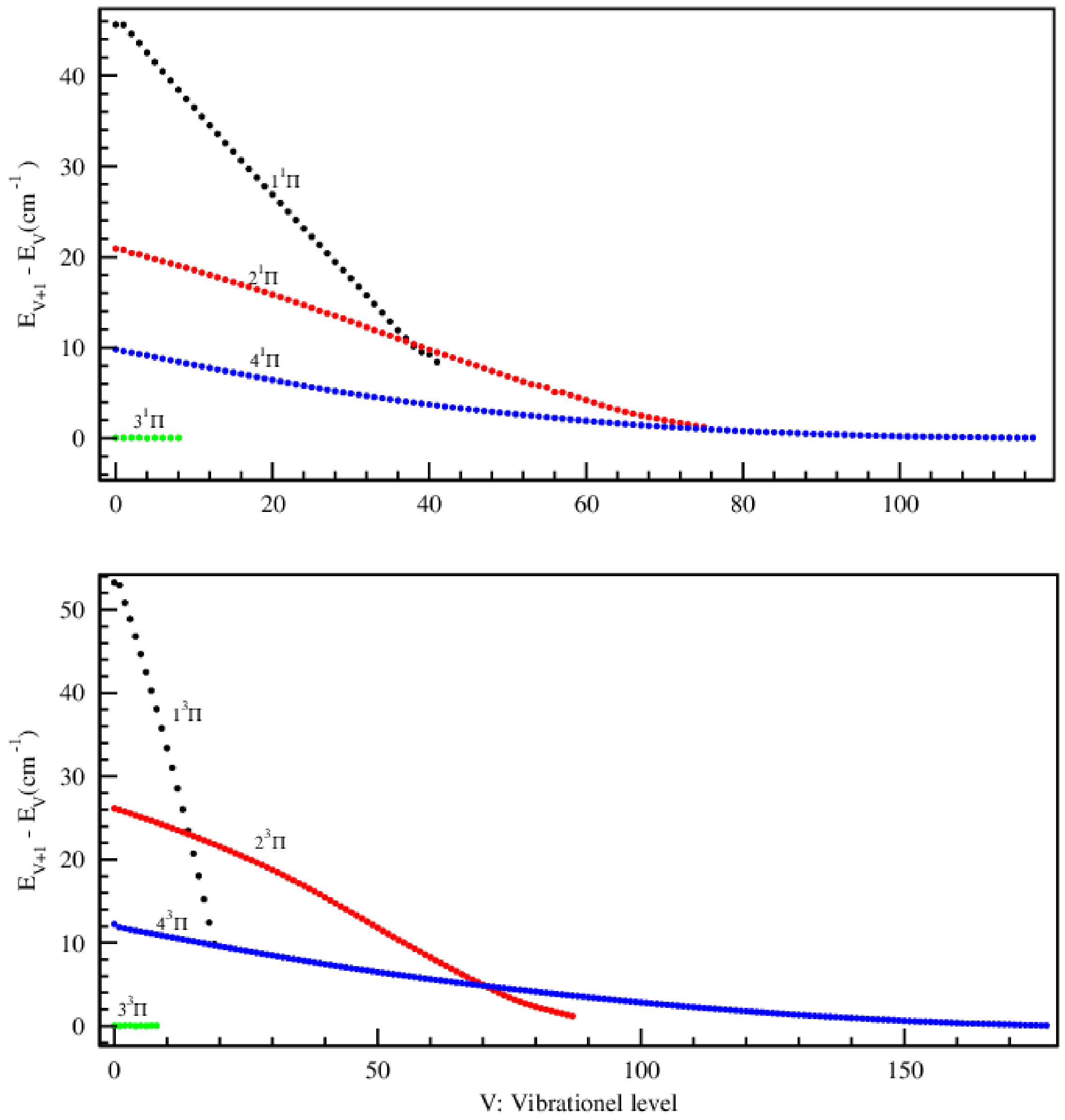

Figure 8. (a) Permanent dipole moment for the $^1\Sigma^+$ states of the $(MgCs)^+$ molecular ion

(b) Zoomed area at distances between 0 and 15 a. u.

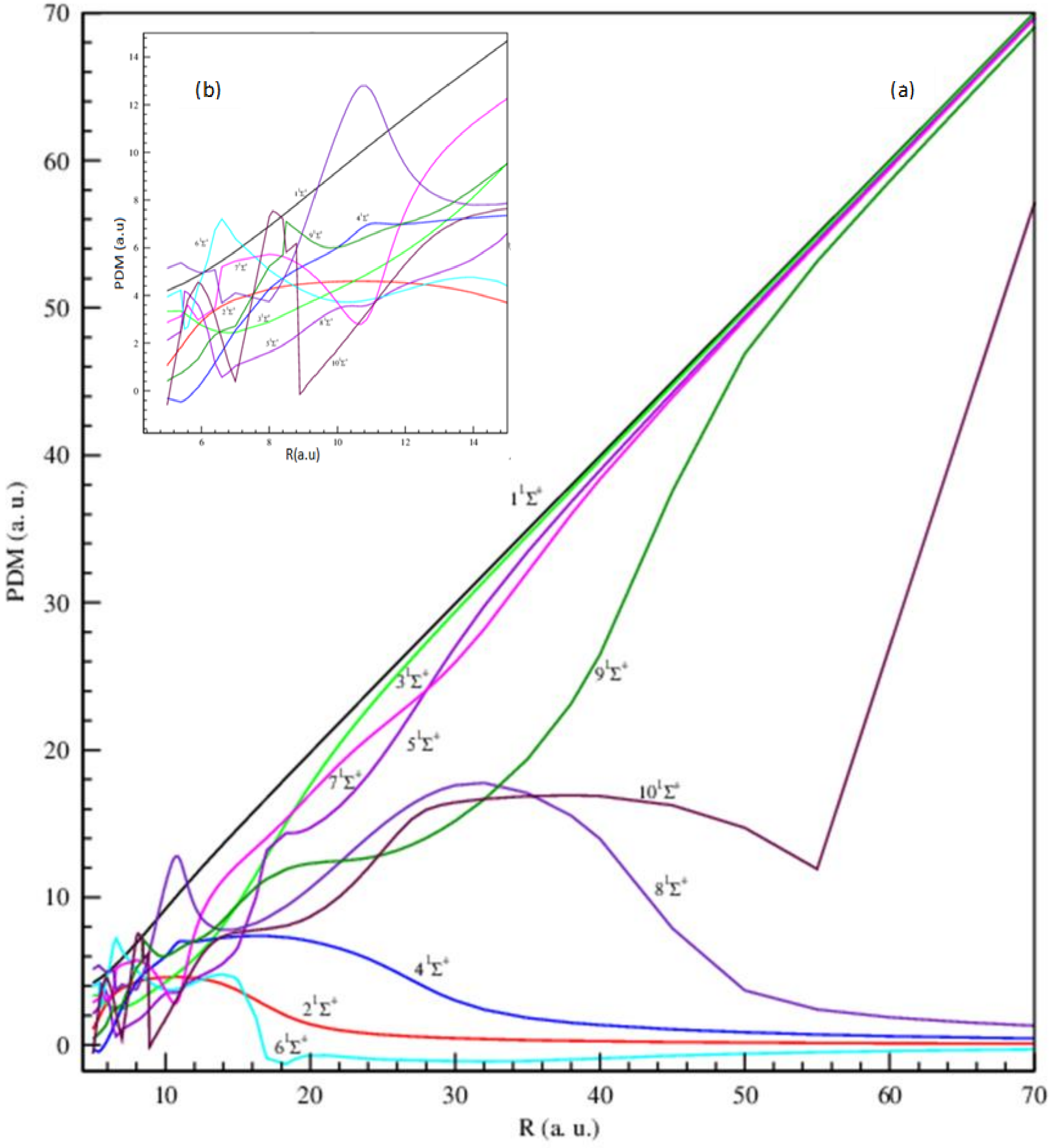

Figure 9. (a) Permanent dipole moment for the $^3\Sigma^+$ states of the $(MgCs)^+$ molecular ion.

(b) Zoomed area at distance between 0 and 15 a. u.

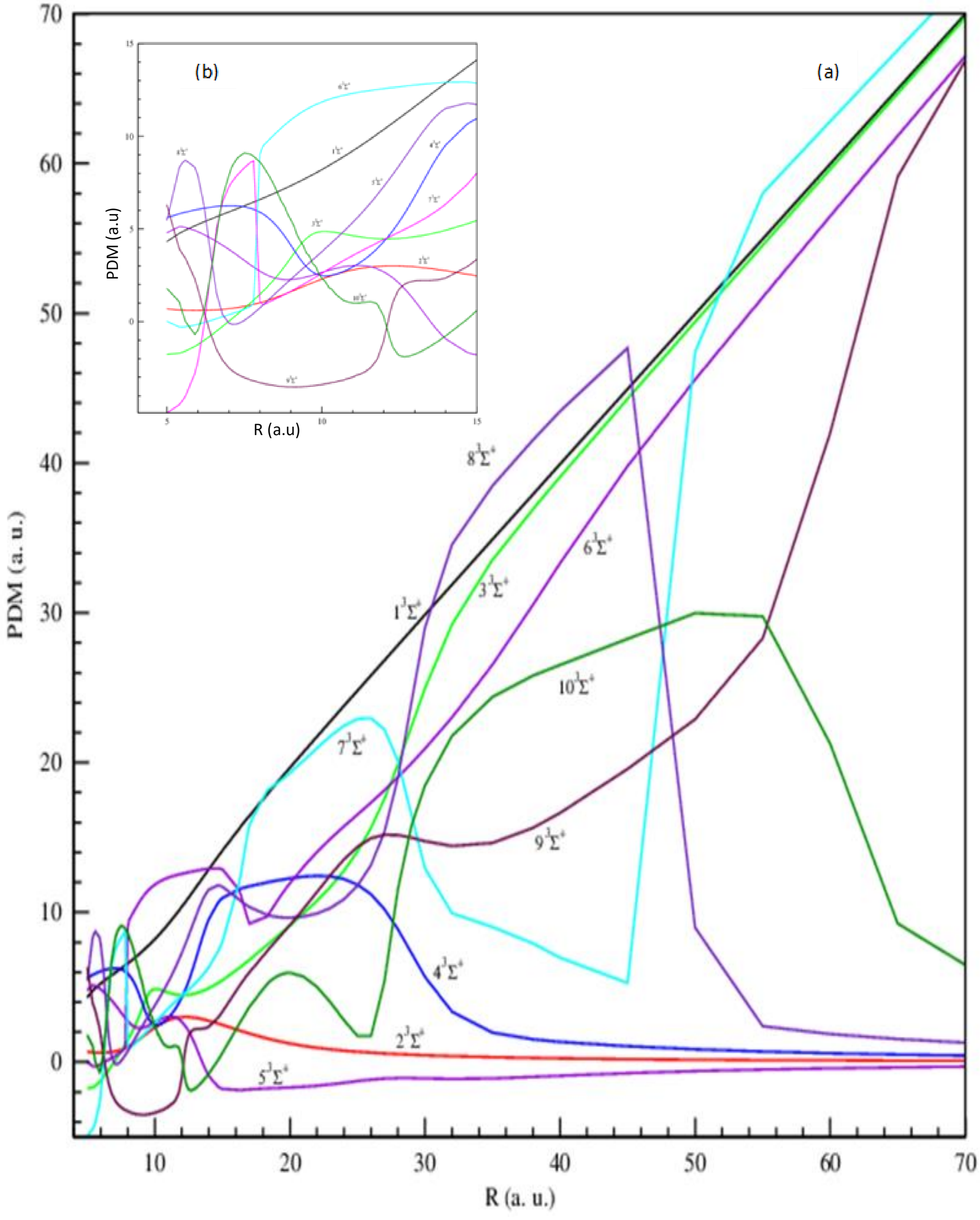

Figure 10. (a) Permanent dipole moment for the $^1\Pi$ states of the $(MgCs)^+$ molecular ion.

(b) Zoomed area at distances between 0 and 15 a. u.

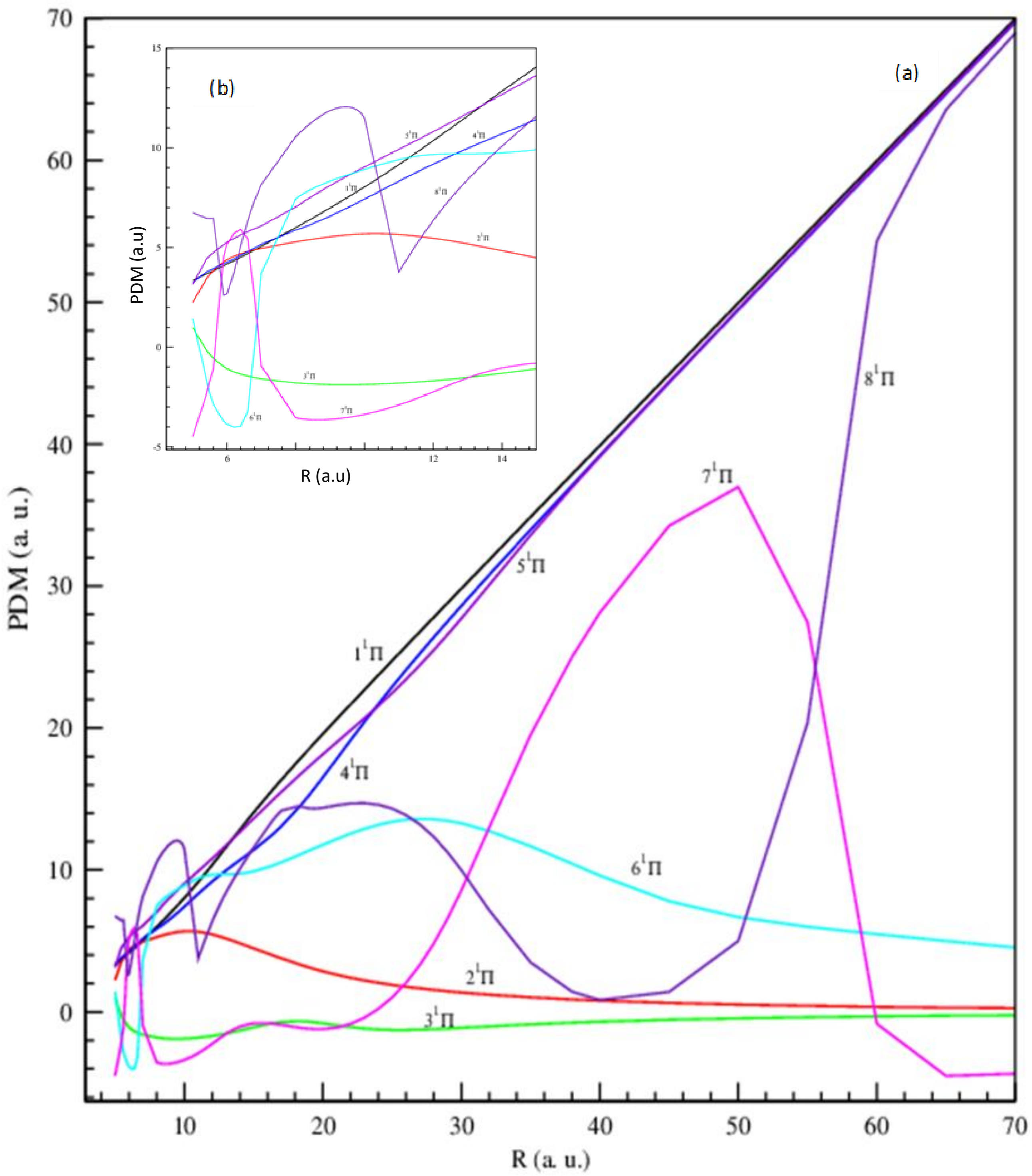

Figure 11. (a) Permanent dipole moment for the $^3\Pi$ states of the (MgCs)+ molecular ion

(b) Zoomed area at distances between 0 and 15 a. u.

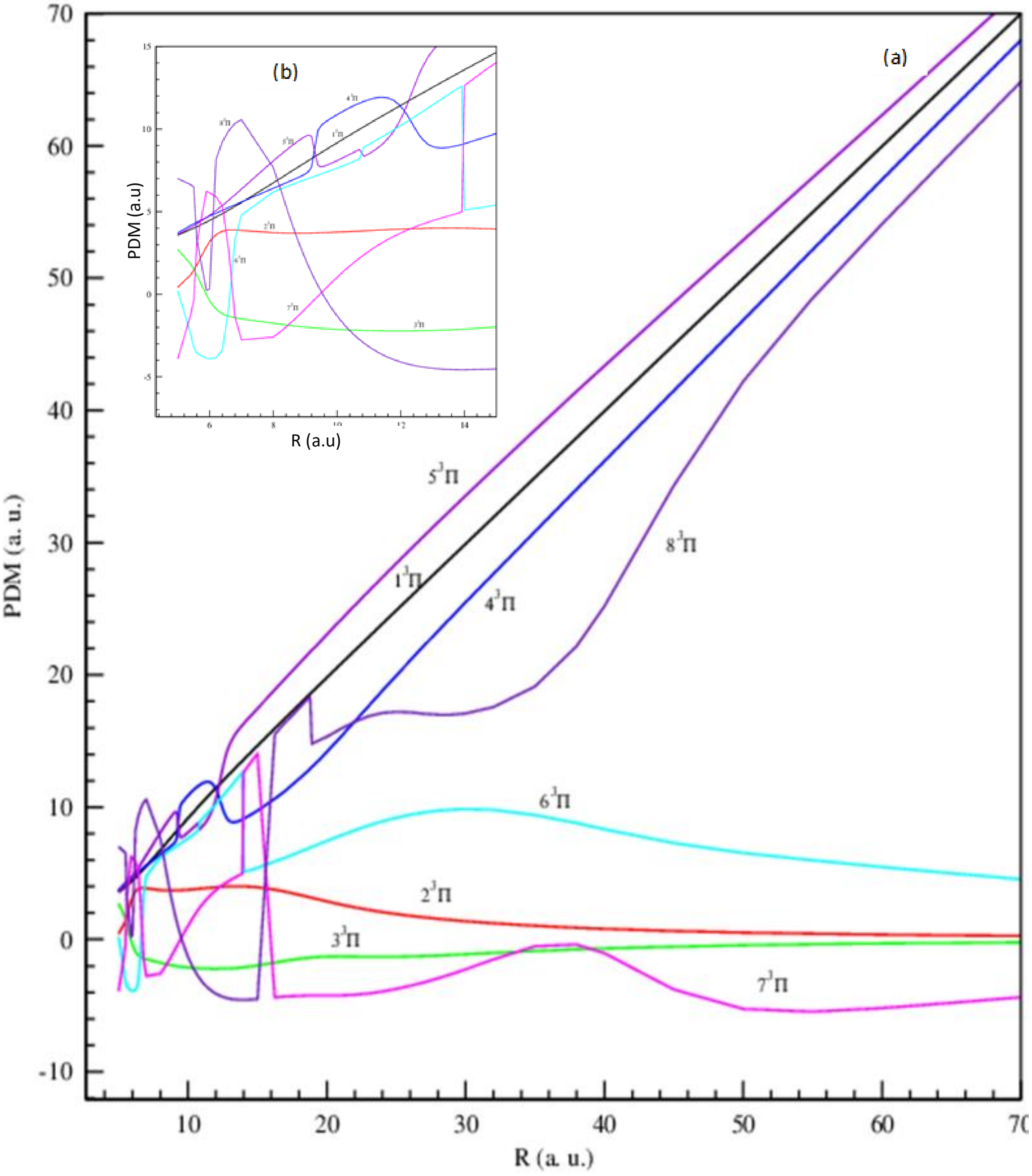

**Figure 12.** Permanent dipole moment for the $^{1,3}\Delta$ states of the $(MgCs)^+$ molecular ion as a function of the inter-nuclear distance $R$.

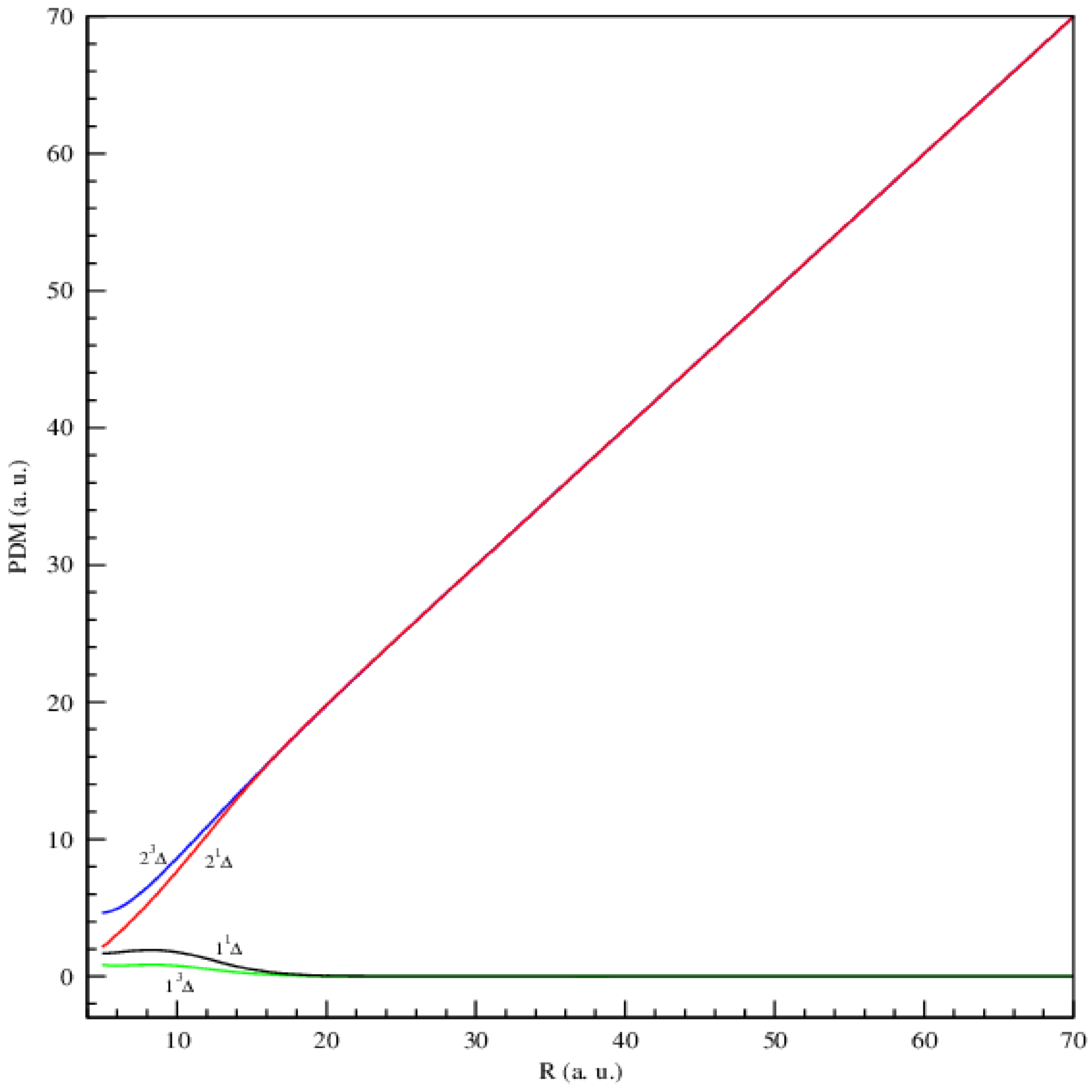

**Figure 13.** Transition dipole moments for selected (1-2, 2-3 and 3-4) $^{1,3}\Sigma^+$ states, as a function of the internuclear distance *R*.

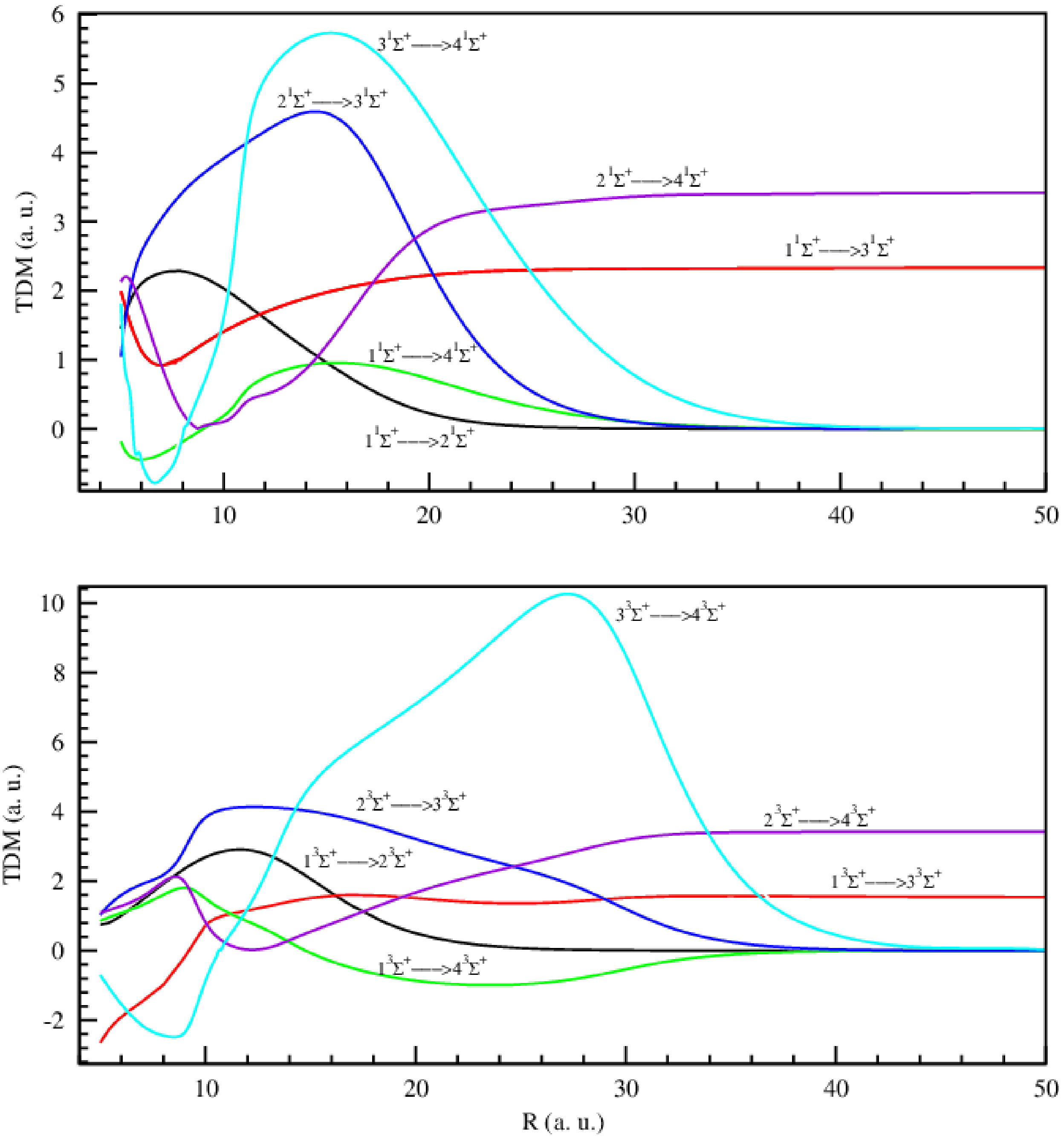

**Figure 14.** Transition dipole moments for selected (1-2, 2-3 and 3-4) $^{1,3}\Pi$ states, as a function of the internuclear distance $R$.

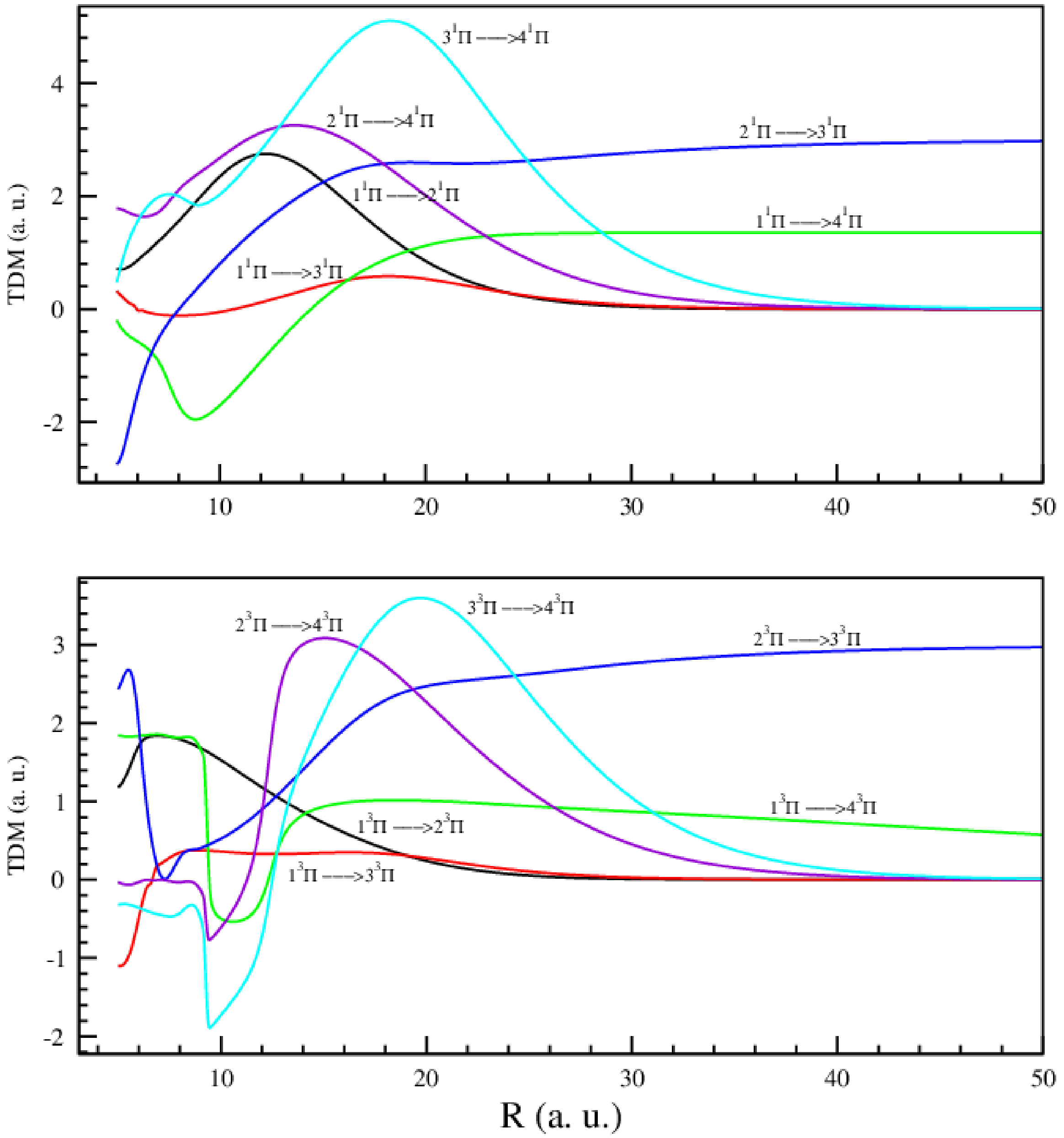

**Figure 15.** Transition dipole moments for selected (1-2, 1-3and 2-3) $^{1,3}\Delta$states, as a function of the internuclear distance *R*.

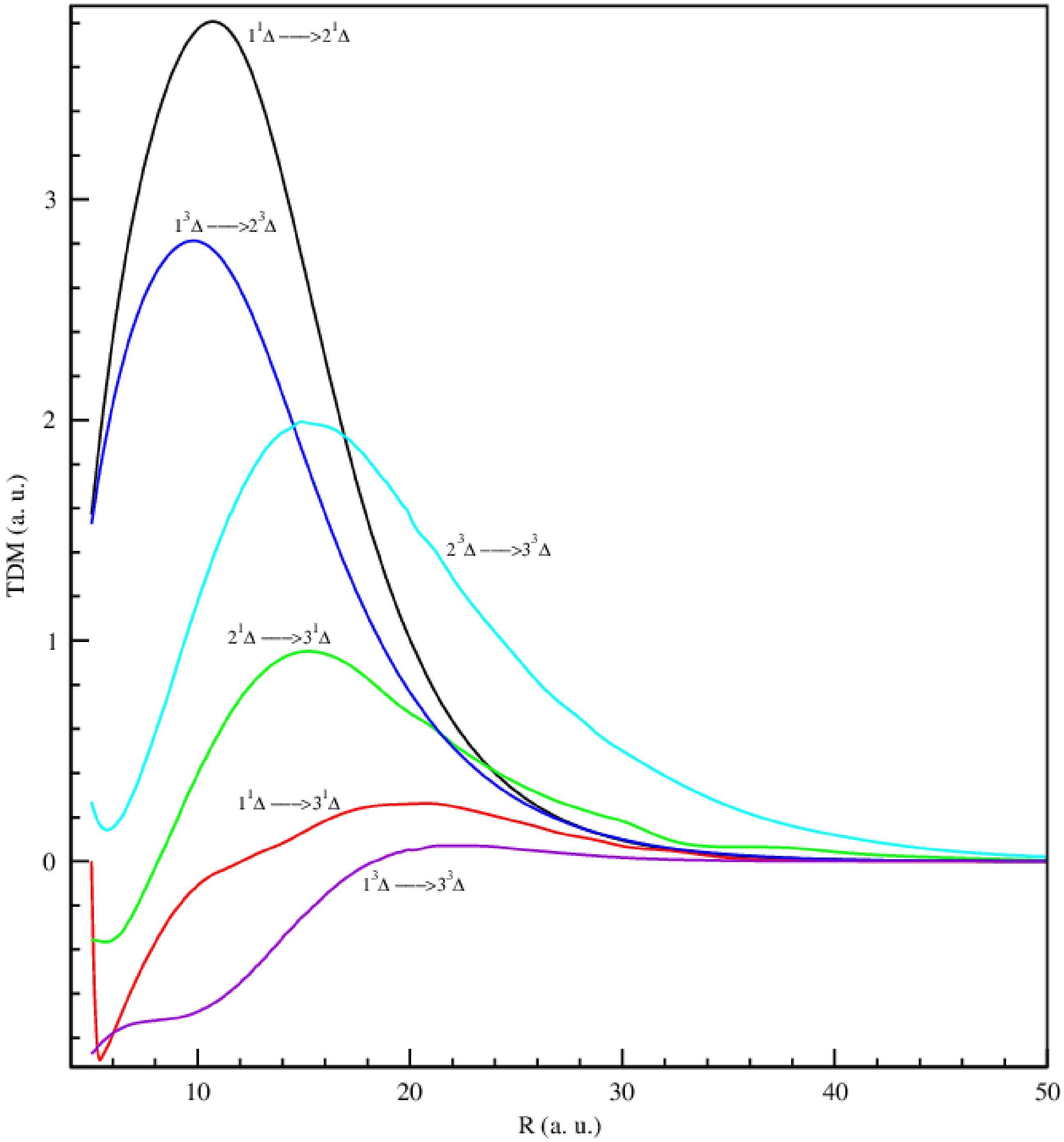

**Figure 16.** .Logarithm of total elastic scattering cross sections ($\sigma_{tot}$) in a.u. as a function of logarithm energy E

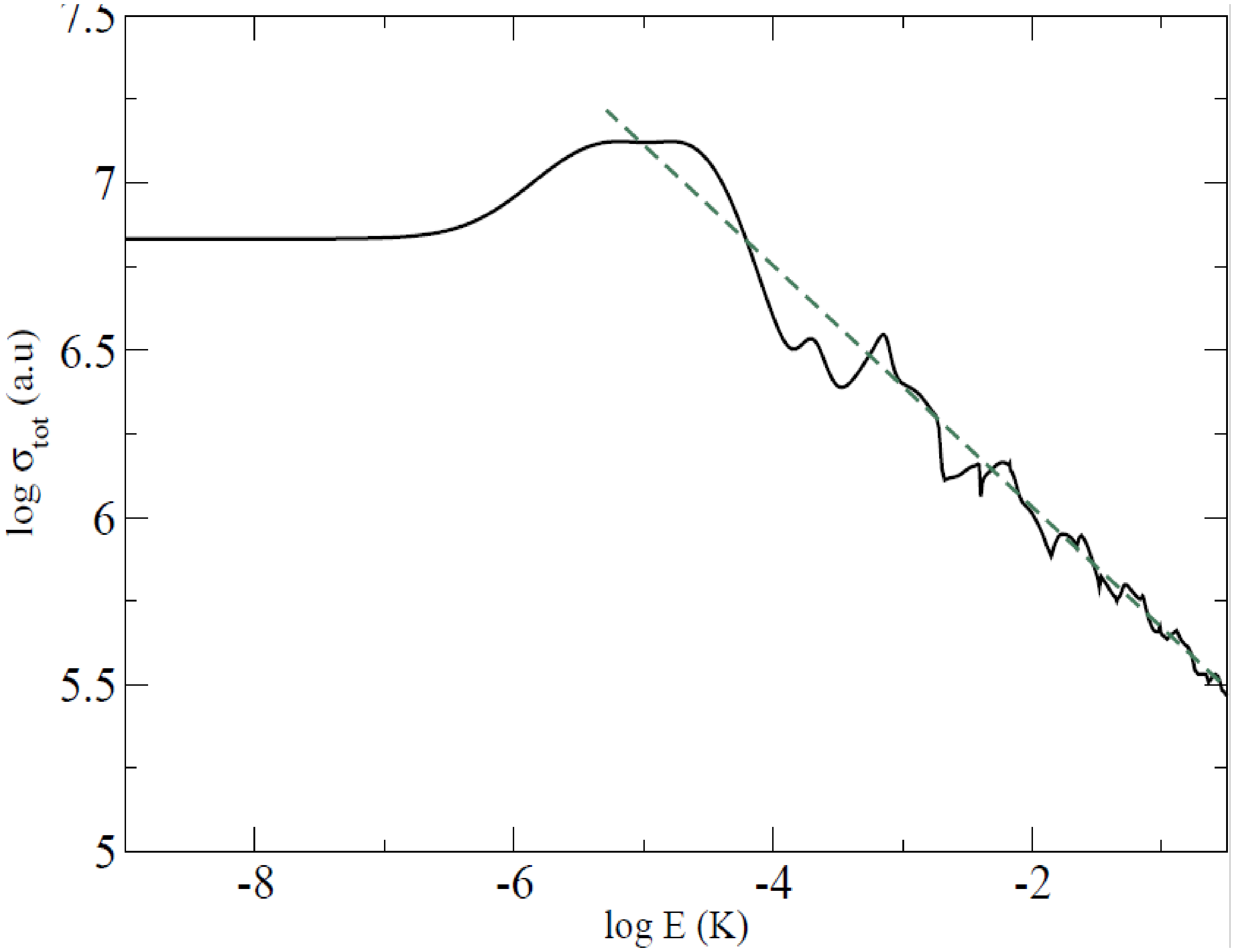

**Figure 17.** The rate constant $K_{PA}$ ($cm^3s^{-1}$) of PA as a function of energy $E$ (in Kelvin) for laser frequency 10 $W/cm^2$ tuned at PA resonance

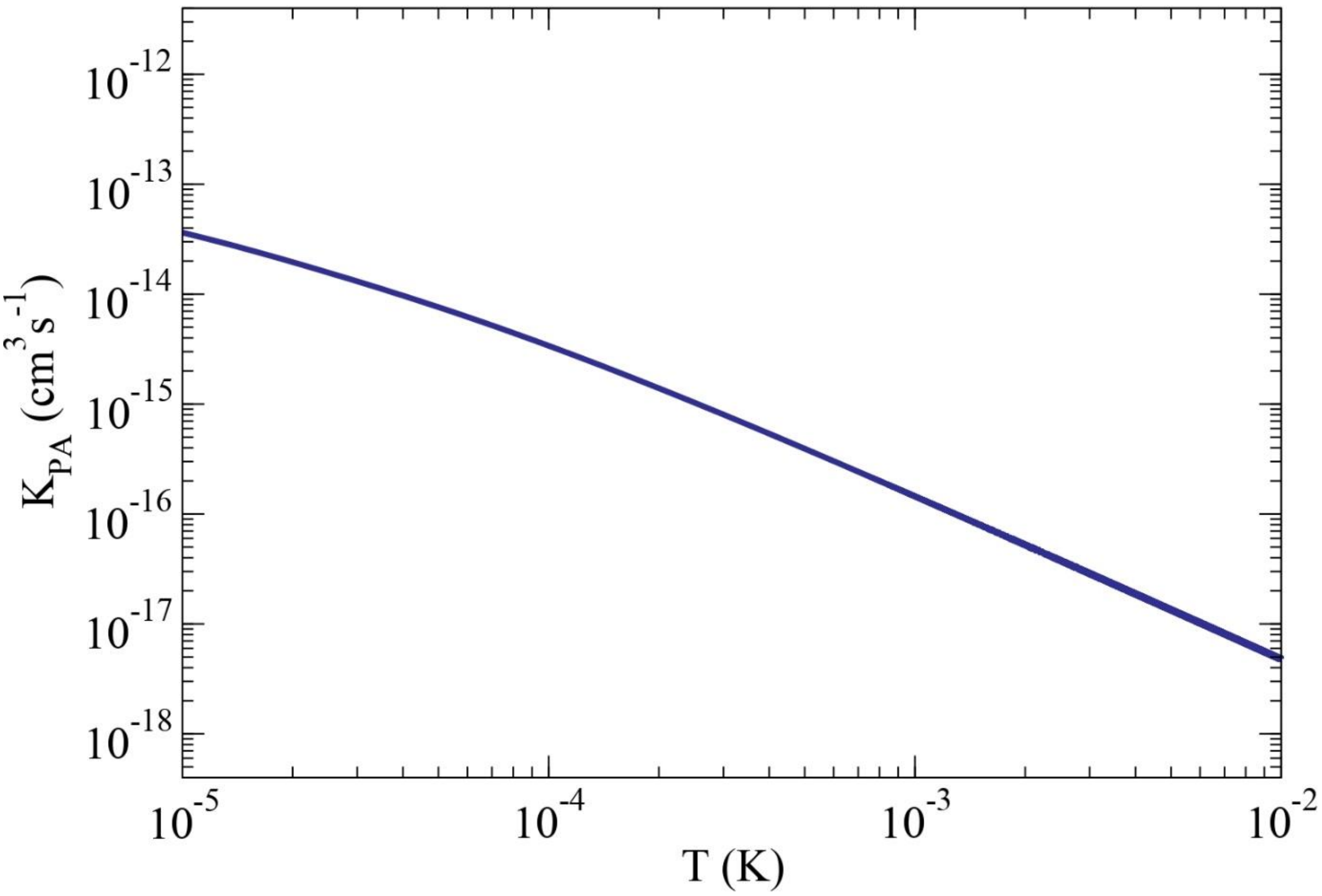

**Figure 18.**The variation of the rate constant of Photoassociation as function of the intensity of the applied laser.

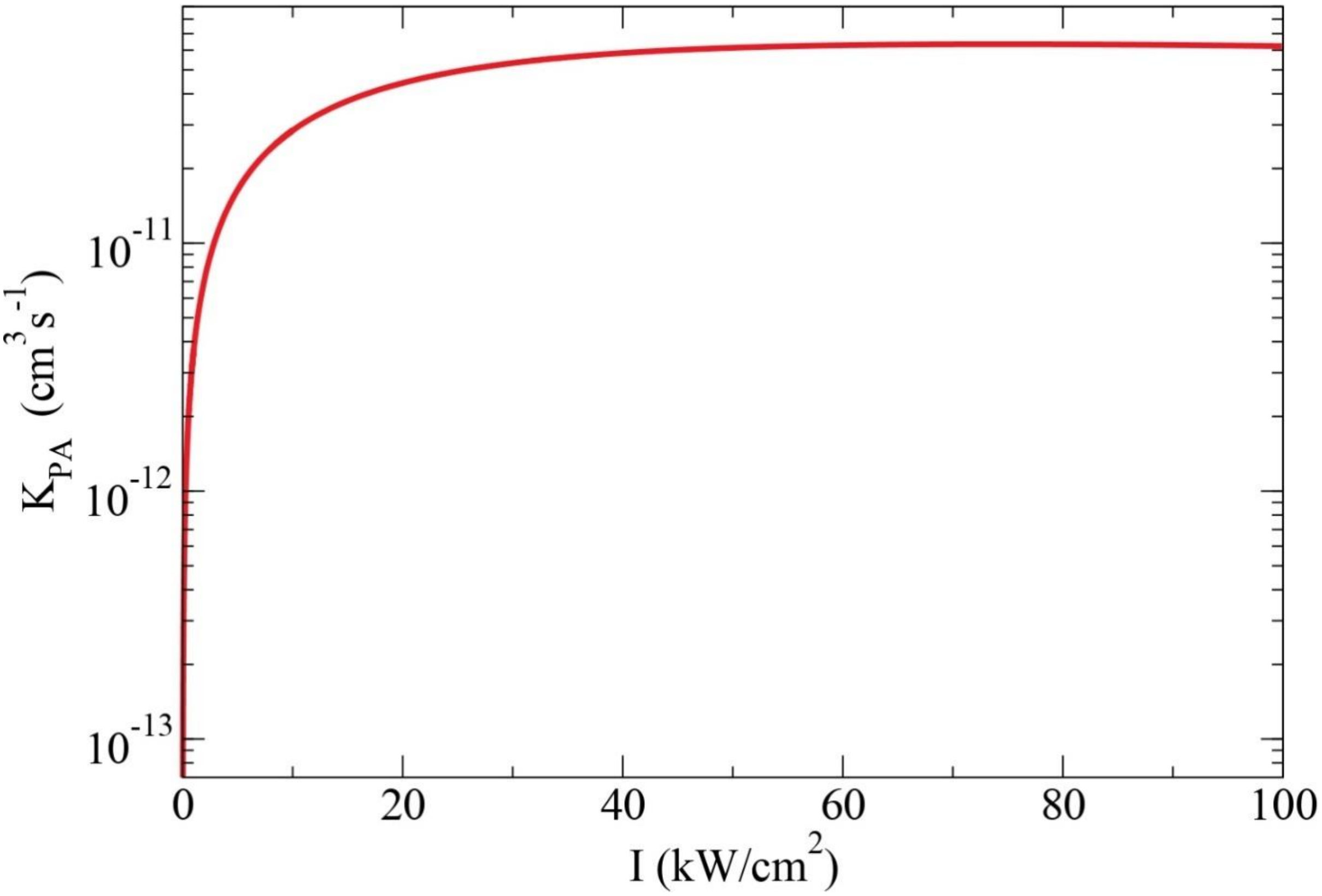